\documentclass[11pt]{JHEP3}
\usepackage{epsf,psfig}
\usepackage{amsfonts,amssymb,latexsym}
\newcommand{\beq}{\begin{eqnarray*}}
\newcommand{\eeq}{\end{eqnarray*}}
\newcommand{\beqn}{\begin{eqnarray}}
\newcommand{\eeqn}{\end{eqnarray}}

\newcommand{\hmbox}[1]{\mbox{\scriptsize #1}}

\newcommand{\Z}{\mathbb{Z}}

\newcommand{\R}{\mathbb{R}}

\newcommand{\F}{\mathbb{F}}
\newcounter{fig}

\newcounter{gif}

\newcommand{\mref}[1]{$(\ref{#1})$}

\def\sqr#1#2{{\vcenter{\vbox{\hrule height.#2pt
            \hbox{\vrule width.#2pt height#1pt \kern#1pt
                  \vrule width.#2pt}\hrule height.#2pt}}}}

\newcommand{\zb}{\bar{z}}

\newcommand{\wb}{\bar{w}}

\newcommand{\mb}{\bar{m}}
\newcommand{\sbar}{\bar{s}}

\newcommand{\hb}{\bar{h}}
\newcommand{\qb}{\bar{q}}

\newcommand{\Eh}{\hat{E}}

\newcommand{\eps}{\epsilon}

\title{On explicit results at the intersection of the $\Z_2$ and
$\Z_4$ orbifold subvarieties in K3 moduli space}
\author{Holger Eberle\\ 
Physikalisches Institut der Universit\"at Bonn, Nu\ss allee 12, 53115 Bonn, Germany\\
E-mail: \email{eberle@th.physik.uni-bonn.de}}

\abstract{We examine the recently found point of intersection between the $\Z_2$ and
$\Z_4$ orbifold subvarieties in the K3 moduli space more closely. First we give an
explicit identification of the coordinates of the respective $\Z_2$ and
$\Z_4$ orbifold theories at this point.
Secondly we construct the explicit identification of conformal field theories
at this point and show the orthogonality of the two subvarieties.}


\preprint{BONN-TH-2004-04}

\begin{document}

\section{Introduction}
Compactifying superstring theory down to six dimensions
we encounter as possibilities for compactification spaces
not only the well understood sixteen dimensional moduli space of
torus compactifications, but also the eighty dimensional
moduli space of quantum K3 surfaces. Although the general mathematical
properties of K3 surfaces are well known (see e.g.\ \cite{BHPV04}),
the precise physical properties of sigma models on these complex surfaces
together with their quantum properties as the B-field
and hence of the respective string vacua are only known
for a nullset of theories in that moduli space, like Gepner or
most of the orbifold models. A lot of pioneering work
on the general structure of the K3 moduli space and the placement
of the above mentioned special models within has been done
in e.g.\ \cite{Sei88,AM94,Asp96,NW99,Wend00,Wend02}.
This paper aims at clarifying two points about the structure of 
the K3 moduli space.

In \cite{NW99,Wend00} it was shown that the subvarieties of 
$\Z_2$ and $\Z_4$ orbifold compactifications in the K3 
moduli space intersect in one point.
But the argument was given in a
rather indirect way. First it made use of the identification
of the quantum surface with its conformal field theory
and performed the identification on the level of the
corresponding conformal field theories. 
The two corresponding $\Z_2$ and $\Z_4$ orbifold conformal
field theories were proven to be equivalent to the
Gepner like model $(\hat{2})^4$, a $\Z_2$ orbifold
of the Gepner model $(2)^4$. 
Secondly it inferred
the identification of two of these three conformal field
theories from the identification of two other specific theories
via the orbifold procedure.
 
In this paper we first give an explicit identification of the 
two lattices and fourplanes corresponding to the two different 
orbifold conformal
field theories at the point of intersection. 
This provides us with a geometric proof 
that the two quantum K3 surfaces are isomorphic.
Secondly we elaborate the explicit identification of the
three conformal field theories at that point and prove
the orthogonality of the two subvarieties of $\Z_2$ and $\Z_4$ 
orbifold compactifications at their point of intersection.
This is a new feature of the intersection which 
has not been clear up to now.
As a byproduct
we see and confirm properties of twistfields in $\Z_4$
orbifold models.

This explicit identification of the $\Z_2$ and $\Z_4$ orbifold conformal
field theories at the point of intersection is important
for further studies on the K3 moduli space. The main open problem
in K3 moduli space is the study of the vast space of yet unknown
sigma models besides the highly symmetric Gepner and orbifold
models. The most promising idea to study these is
to use conformal deformation theory (see e.g.\ \cite{Kad78}) starting from known
theories. This should be easiest on the nontrivial geodesics
spanned between the $\Z_2$ and $\Z_4$ orbifold subvarieties
as these specific geodesics have known starting and endpoints.
But for this endeavour we crucially need the exact relation
between the coordinates on the two subvarieties given
by the explicit identification in this article.

Furthermore, the orthogonality of the two subvarieties proven
in this article allows to describe some of the nontrivial 
twistfield deformations of one type of orbifold by the well known
torus type deformations of the other, at the point of intersection.
This makes it possible to test and to understand more about
these nontrivial deformations which are usually much harder to compute.
One open problem to possibly analyse this way is
the curious symmetry which we observed \cite{Ebe01}
in the dependence of the deformation
of a vertex operator in an orbifold model on the conformal dimension
about the point $h=1/8$.



\section{Lattices of orbifold theories\label{section2}}

Let us first recall some facts about the structure of lattices signifying certain
orbifold theories in K3 moduli space. We try to stick to the conventions of 
\cite{NW99,Wend00,Wend02} where a much more detailed presentation is given.

The general structure of the K3 moduli space is given by
\beq
\mbox{O}^+ (\Gamma^{(4,20)}) \backslash \mbox{O}^+ (4,20) / \mbox{SO}(4) \times \mbox{O}(20)
& \cong & \mbox{O}^+ (\Gamma^{(4,20)}) \backslash \mbox{O} (4,20) / \mbox{O}(4) \times \mbox{O}(20)
\eeq
where $\mbox{O}^+ (\Gamma^{(4,20)})$ signifies the discrete duality group. As we are only interested
in local properties in this paper, we will only deal with the unique smooth, simply connected covering 
space $\mbox{O} (4,20) / \mbox{O}(4) \times \mbox{O}(20)$. This is a Grassmanian
space whose points can be described by a fourplane in an even selfdual  $24$-dimensional lattice
$\Gamma^{(4,20)}$ of signature $(4,20)$. 
Such even selfdual lattices are known to be
unique up to isometries for a given signature $(m,n)$, if $m>0$ and $n>0$
(see e.g.\ \cite{Asp96}).
\cite{NW99,Wend00} deduce how to embed the torus moduli space into the
K3 moduli space via the $\Z_n$, $n=2,3,4,6$, orbifold procedure and thus give
the explicit coordinates, i.e.\ fourplanes in corresponding lattices, for the
points of these orbifold subvarieties in the K3 moduli space. In this parametrisation
the reference lattices are given by the even $\Z$ cohomology of that surface.
As we only need the
coordinates of two special models at the point of intersection of the 
$\Z_2$ and $\Z_4$ orbifold subvarieties, we will only quote the lattices for
$\Z_2$ and $\Z_4$ orbifold models in the following as well as the fourplanes and their
orthogonal span for the two special models we consider.

\subsection{Reference lattices for $\Z_2$ and $\Z_4$ orbifold models\label{lattices}}
Let ${\cal T} (\Lambda,B)$ signify the torus theory with torus lattice $\Lambda$ and B-field $B$
and let $\mu_i$ be the dual vectors to the generators $\lambda_i$ of $\Lambda$.
From any torus we can construct a K3 surface by using the orbifolding technique
and blowing up the orbifold singularities. If we take the metric in the orbifold limit
we will call this an orbifold surface. 
For any such orbifold surface $X$, let $\upsilon^{\circ}  \in {\cal H}^0 \; (X , \Z)$ and 
$\upsilon \in {\cal H}^4 \; (X , \Z)$ be the
generators of the respective one-dimensional cohomology groups, with 
$\langle \upsilon^{\circ}, \upsilon \rangle = 1$, $||\upsilon^{\circ}||^2=||\upsilon||^2 = 0$.
$E_i \in {\cal H}^2 \; (X , \Z)$ signify the 
exceptional divisors of norm $||E_i||^2=-2$ obtained
by blowing up the orbifold singularities of the surface. Also let 
$P_{jk} := \mbox{span}_{\F_2} \, (f_j,f_k)$ with $j,k \in \{1,\dots,4\}$
and $f_j$ the $j$th standard basis vector. 

Let ${\cal K} (\Lambda,B)$ signify the $\Z_2$ orbifold of ${\cal T} (\Lambda,B)$.
The orbifold group acts via multiplication by $-1$ on $\R^4$.
The lattice of even cohomology $\Gamma = {\cal H}^{\hmbox{even}} \; (X , \Z)$ of the K3 surface $X$
corresponding to ${\cal K} (\Lambda,B)$ is generated by (\cite{NW99,Wend00,Wend02})
\beqn \label{Z2_eq1}
\hat{\upsilon} &=& \sqrt{2} \; \upsilon \; , \nonumber \\
\hat{\upsilon}^{\circ} &=& \frac{1}{\sqrt{2}} \upsilon^{\circ} - \frac{1}{4} \sum_{i\in I} \, E_i + \sqrt{2} \; \upsilon
\eeqn
and the sublattice $\hat{\Gamma}_{\Z_2}$ (using $\hat{E}_i := - \frac{1}{\sqrt{2}} \upsilon + E_i$)
\beqn \label{Z2_eq2}
\hat{\Gamma_{\Z_2}} := 
\mbox{span}_{\Z} \; \left(\frac{1}{\sqrt{2}} \, \mu_j \wedge \mu_k + \frac{1}{2} \sum_{i \in P_{jk}} \, \hat{E}_{i+l} 
\;, \; \hat{E}_m ; \; l, m \in I \right) \; .
\eeqn
$I = (\F_2)^4$ parametrises the sixteen $\Z_2$ orbifold fixpoints which have been blown up 
with one exceptional divisor each.

Similarly let $\Z_4 (\Lambda,B)$ signify the $\Z_4$ orbifold of ${\cal T} (\Lambda,B)$, where
the orbifold group acts like $\mu_1 \mapsto \mu_2$, $\mu_2 \mapsto -\mu_1$, $\mu_3 \mapsto -\mu_4$,
$\mu_4 \mapsto \mu_3$.
The lattice of the even cohomology $\Gamma = {\cal H}^{\hmbox{even}} \; (Y , \Z)$ of the corresponding 
K3 surface $Y$ is generated by (\cite{NW99,Wend00,Wend02})\footnote{We 
apologise for the abuse of notation
using the same symbols $\hat{E}$ etc.\ as in the ${\cal K} (\Lambda,B)$ models in order to prevent
a proliferation of indices.
We hope that a distinction between the different models is clearly visible from the context.}
\beq
\hat{\upsilon} &=& 2 \; \upsilon \; , \nonumber \\
\hat{\upsilon}^{\circ} &=& \frac{1}{2} \upsilon^{\circ} - \frac{1}{4} \sum_{i\in I^{(2)}} \, E_i 
- \frac{1}{8} \sum_{i\in I^{(4)}} \, (3 E^{(+)}_i + 4 E^{(0)}_i + 3 E^{(-)}_i)  + 2 \; \upsilon
\eeq
and the sublattice $\hat{\Gamma}_{\Z_4}$ spanned by 
(using $\hat{E}_i := - \langle E_i, \hat{\upsilon}^{\circ} \rangle \, \hat{\upsilon} + E_i$)
\beq
&& \frac{1}{2} \, \mu_1 \wedge \mu_2 + \frac{1}{2} \, \Eh_{(0,0,1,0)+\eps (1,1,0,0)} 
+ \frac{1}{4} \sum_{i\in P_{34} \cap I^{(4)}} \Eh_{i+\eps (1,1,0,0)} \quad \mbox{with } \eps \in \{0,1\} \; , \nonumber \\
&& \frac{1}{2} \, \mu_3 \wedge \mu_4 - \frac{1}{2} \, \Eh_{(1,0,0,0)+\eps (0,0,1,1)} 
- \frac{1}{4} \sum_{i\in P_{12} \cap I^{(4)}} \Eh_{i+\eps (0,0,1,1)} \quad \mbox{with } \eps \in \{0,1\} \; , \nonumber \\
&& \frac{1}{2} \, (\mu_1 \wedge \mu_3 + \mu_4 \wedge \mu_2) - \frac{1}{2} \sum_{i\in P_{13}} \Eh_{i+j} + \Eh_j 
\quad \mbox{with } j \in I^{(4)} \; , \nonumber \\
&& \frac{1}{2} \, (\mu_1 \wedge \mu_4 + \mu_2 \wedge \mu_3) - \frac{1}{2} \sum_{i\in P_{14}} \Eh_{i+j} + \Eh_j 
\quad \mbox{with } j \in I^{(4)} \; , \nonumber \\
&& \Eh_k \quad \mbox{with }  k \in I^{(2)} \cup I^{(4)} \; .
\eeq
$I^{(4)} = \{0000,0011,1100,1111\}$ parametrises the four $\Z_4$ orbifold fixpoints 
whose blow-up produces three exceptional divisors each, 
$I^{(2)} = \{0100,0001,0111,1101,0110,0101\}$ the six $\Z_2$ orbifold fixpoints.

\subsection{The $\Z_2$ orbifold ${\cal K} (\Z^4,0)$}
In the special model ${\cal K} (\Z^4,0)$,
the positive definite fourplane representing this point in
moduli space is spanned by the following pairwise orthogonal lattice vectors of norm $||A_i||=4$ 
(within the above described lattice of signature $(4,20)$ for $\Z_2$ orbifold surfaces)
\beq
A_{21} &=& \sqrt{2} \; (e_1 \wedge e_2 + e_3 \wedge e_4) \nonumber \\
A_{22} &=& \sqrt{2} \; (e_1 \wedge e_3 + e_4 \wedge e_2) \nonumber \\
A_{23} &=& \sqrt{2} \; (e_1 \wedge e_4 + e_2 \wedge e_3) \nonumber \\
A_{24} &=& 2\hat{\upsilon}^{\circ} + \frac{1}{2} \sum_{i \in I^{(2)}} \hat{E}_i + \,3 \, \hat{\upsilon} \; .
\eeq
The space orthogonal to this fourplane is likewise spanned by the following pairwise orthogonal
lattice vectors of norm $||A_i||=-4$
\beq
A_{1} &=& \sqrt{2} \; (e_1 \wedge e_2 - e_3 \wedge e_4) \nonumber \\
A_{2} &=& \sqrt{2} \; (e_1 \wedge e_3 - e_4 \wedge e_2) \nonumber \\
A_{3} &=& \sqrt{2} \; (e_1 \wedge e_4 - e_2 \wedge e_3) \nonumber \\
A_{4} &=& 2 \, \hat{\upsilon}^{\circ} + \frac{1}{2} \sum_{i \in I^{(2)}} \hat{E}_i + \, \hat{\upsilon} \nonumber \\
A_{5} &=& \frac{1}{2} (\hat{E}_{0000} - \hat{E}_{1100} - \hat{E}_{0011} + \hat{E}_{1111} - \hat{E}_{1010} 
- \hat{E}_{1001} - \hat{E}_{0110} - \hat{E}_{0101}) -  \, \hat{\upsilon} \nonumber \\
A_{6} &=& \frac{1}{2} (\hat{E}_{0000} - \hat{E}_{1100} - \hat{E}_{0011} + \hat{E}_{1111} + \hat{E}_{1010} 
+ \hat{E}_{1001} + \hat{E}_{0110} + \hat{E}_{0101}) + \, \hat{\upsilon} \nonumber \\
A_{7} &=& \frac{1}{2} (\hat{E}_{0000} + \hat{E}_{1100} + \hat{E}_{0011} + \hat{E}_{1111} - \hat{E}_{1010} 
+ \hat{E}_{1001} + \hat{E}_{0110} - \hat{E}_{0101}) + \, \hat{\upsilon} \nonumber \\
A_{8} &=& \frac{1}{2} (\hat{E}_{0000} + \hat{E}_{1100} + \hat{E}_{0011} + \hat{E}_{1111} + \hat{E}_{1010} 
- \hat{E}_{1001} - \hat{E}_{0110} + \hat{E}_{0101}) + \, \hat{\upsilon} \nonumber \\
A_{9} &=& \frac{1}{2} (\hat{E}_{1000} + \hat{E}_{0100} + \hat{E}_{0010} + \hat{E}_{0001} - \hat{E}_{1110} 
- \hat{E}_{1101} - \hat{E}_{1011} - \hat{E}_{0111}) \nonumber \\
A_{10} &=& \frac{1}{2} (\hat{E}_{1000} + \hat{E}_{0100} - \hat{E}_{0010} - \hat{E}_{0001} + \hat{E}_{1110} 
+ \hat{E}_{1101} - \hat{E}_{1011} - \hat{E}_{0111}) \nonumber \\
A_{11} &=& \frac{1}{2} (\hat{E}_{1000} - \hat{E}_{0100} + \hat{E}_{0010} - \hat{E}_{0001} + \hat{E}_{1110} 
- \hat{E}_{1101} + \hat{E}_{1011} - \hat{E}_{0111}) \nonumber \\
A_{12} &=& \frac{1}{2} (\hat{E}_{1000} - \hat{E}_{0100} - \hat{E}_{0010} + \hat{E}_{0001} - \hat{E}_{1110} 
+ \hat{E}_{1101} + \hat{E}_{1011} - \hat{E}_{0111}) \nonumber \\
A_{13} &=& \frac{1}{2} (\hat{E}_{0000} - \hat{E}_{1100} + \hat{E}_{0011} - \hat{E}_{1111} - \hat{E}_{1010} 
+ \hat{E}_{1001} - \hat{E}_{0110} + \hat{E}_{0101}) \nonumber \\
A_{14} &=& \frac{1}{2} (\hat{E}_{0000} - \hat{E}_{1100} + \hat{E}_{0011} - \hat{E}_{1111} + \hat{E}_{1010} 
- \hat{E}_{1001} + \hat{E}_{0110} - \hat{E}_{0101}) \nonumber \\
A_{15} &=& \frac{1}{2} (\hat{E}_{0000} + \hat{E}_{1100} - \hat{E}_{0011} - \hat{E}_{1111} - \hat{E}_{1010} 
- \hat{E}_{1001} + \hat{E}_{0110} + \hat{E}_{0101}) \nonumber \\
A_{16} &=& \frac{1}{2} (\hat{E}_{0000} + \hat{E}_{1100} - \hat{E}_{0011} - \hat{E}_{1111} + \hat{E}_{1010} 
+ \hat{E}_{1001} - \hat{E}_{0110} - \hat{E}_{0101}) \nonumber \\
A_{17} &=& \frac{1}{2} (\hat{E}_{1000} + \hat{E}_{0100} + \hat{E}_{0010} - \hat{E}_{0001} - \hat{E}_{1110} 
+ \hat{E}_{1101} + \hat{E}_{1011} + \hat{E}_{0111}) + \, \hat{\upsilon} \nonumber \\
A_{18} &=& \frac{1}{2} (\hat{E}_{1000} + \hat{E}_{0100} - \hat{E}_{0010} + \hat{E}_{0001} + \hat{E}_{1110} 
- \hat{E}_{1101} + \hat{E}_{1011} + \hat{E}_{0111}) + \, \hat{\upsilon} \nonumber \\
A_{19} &=& \frac{1}{2} (\hat{E}_{1000} - \hat{E}_{0100} + \hat{E}_{0010} + \hat{E}_{0001} + \hat{E}_{1110} 
+ \hat{E}_{1101} - \hat{E}_{1011} + \hat{E}_{0111}) + \, \hat{\upsilon} \nonumber \\
A_{20} &=& \frac{1}{2} (\hat{E}_{1000} - \hat{E}_{0100} - \hat{E}_{0010} - \hat{E}_{0001} - \hat{E}_{1110} 
- \hat{E}_{1101} - \hat{E}_{1011} + \hat{E}_{0111}) - \, \hat{\upsilon} \; .
\eeq

\subsection{The $\Z_4$ orbifold $\Z_4 (\frac{1}{\sqrt{2}} D_4, B^*)$}
Now we want to look at the special theory $\Z_4 (\frac{1}{\sqrt{2}} D_4, B^*)$,
with the lattice $D_4 = \{ x \in \Z_4 | \sum x_i \equiv 0 \; (2) \}$.
The $B$ field in this theory is given as the 
$\Lambda^* \otimes \R \rightarrow \Lambda \otimes \R$ map (\cite{NW99,Wend00})
\beqn
\label{Bfield}
B^* &=&
\left( \begin{array}{cccc}
0 & 1 & 0 & 0 \\
-1 & 0 & 0 & 0 \\
0 & 0 & 0 & 1 \\
0 & 0 & -1 & 0
\end{array} \right) \; .
\eeqn
In this model the positive definite fourplane representing this point in
moduli space is spanned by the following pairwise orthogonal lattice vectors of norm $||B_i||=4$ 
(within the above described lattice of signature $(4,20)$ for $\Z_4$ orbifold surfaces)
\beq
B_{21} &=& 2 \,  e_1 \wedge e_2 + (e_1 \wedge e_3 + e_4 \wedge e_2) - (e_1 \wedge e_4 + e_2 \wedge e_3) - \, \hat{\upsilon} \nonumber \\
B_{22} &=& (e_1 \wedge e_3 + e_4 \wedge e_2) + (e_1 \wedge e_4 + e_2 \wedge e_3) \nonumber \\
B_{23} &=& 2 \,  e_3 \wedge e_4 - (e_1 \wedge e_3 + e_4 \wedge e_2) + (e_1 \wedge e_4 + e_2 \wedge e_3) \nonumber \\
B_{24} &=& 4 \, \hat{\upsilon}^{\circ} + 2 \,  e_1 \wedge e_2 
+ (e_1 \wedge e_3 + e_4 \wedge e_2) - (e_1 \wedge e_4 + e_2 \wedge e_3) \nonumber \\
&& \quad + \sum_{i \in I^{(2)}} \hat{E}_i +  \frac{1}{2} \sum_{i \in I^{(4)}} 
(3 \hat{E}_i^{(+)} + 4 \hat{E}_i^{(0)} + 3 \hat{E}_i^{(-)}) + \, 4 \, \hat{\upsilon} \; .
\eeq
The space orthogonal to this fourplane is likewise spanned by the following pairwise orthogonal 
lattice vectors of norm $||B_i||=-4$
\beq
B_{1} &=& 2 \,  (e_1 \wedge e_2 - e_3 \wedge e_4) + (e_1 \wedge e_3 + e_4 \wedge e_2) - (e_1 \wedge e_4 + e_2 \wedge e_3) \nonumber \\
B_{2} &=&  4 \, \hat{\upsilon}^{\circ} + 2 \,  e_1 \wedge e_2 
+ (e_1 \wedge e_3 + e_4 \wedge e_2) - (e_1 \wedge e_4 + e_2 \wedge e_3) \nonumber \\
&& \quad + \sum_{i \in I^{(2)}} \hat{E}_i +  \frac{1}{2} \sum_{i \in I^{(4)}} 
(3 \hat{E}_i^{(+)} + 4 \hat{E}_i^{(0)} + 3 \hat{E}_i^{(-)}) + \, 3 \, \hat{\upsilon} \nonumber \\
B_{3} &=& \hat{E}_{0100} - \hat{E}_{0111} \nonumber \\
B_{4} &=& \hat{E}_{0001} - \hat{E}_{1101} \nonumber \\
B_{5} &=& \hat{E}_{0110} - \hat{E}_{1010} \nonumber \\
B_{6} &=& \hat{E}_{0100} + \hat{E}_{0111} + \, \hat{\upsilon} \nonumber \\
B_{7} &=& \hat{E}_{0001} + \hat{E}_{1101} + \, \hat{\upsilon} \nonumber \\
B_{8} &=& \hat{E}_{0110} + \hat{E}_{1010} + \, \hat{\upsilon} \nonumber \\
B_{9} &=& \frac{1}{2} (\hat{E}_{0000}^{(+)} + 2\hat{E}_{0000}^{(0)} + \hat{E}_{0000}^{(-)} 
+ \hat{E}_{1100}^{(+)} + 2\hat{E}_{1100}^{(0)} + \hat{E}_{1100}^{(-)} + \hat{E}_{0011}^{(+)} - \hat{E}_{0011}^{(-)}  \nonumber \\
&& \quad - \hat{E}_{1111}^{(+)} + \hat{E}_{1111}^{(-)}) + \, \hat{\upsilon} \nonumber \\
B_{10} &=& \frac{1}{2} (\hat{E}_{0000}^{(+)} + 2\hat{E}_{0000}^{(0)} + \hat{E}_{0000}^{(-)} 
+ \hat{E}_{1100}^{(+)} + 2\hat{E}_{1100}^{(0)} + \hat{E}_{1100}^{(-)} - \hat{E}_{0011}^{(+)} + \hat{E}_{0011}^{(-)} \nonumber \\
&& \quad + \hat{E}_{1111}^{(+)} - \hat{E}_{1111}^{(-)}) + \, \hat{\upsilon} \nonumber \\
B_{11} &=& \frac{1}{2} (\hat{E}_{0000}^{(+)} + 2\hat{E}_{0000}^{(0)} + \hat{E}_{0000}^{(-)} 
- \hat{E}_{1100}^{(+)} - 2\hat{E}_{1100}^{(0)} - \hat{E}_{1100}^{(-)} - \hat{E}_{0011}^{(+)} + \hat{E}_{0011}^{(-)} \nonumber \\
&& \quad - \hat{E}_{1111}^{(+)} + \hat{E}_{1111}^{(-)}) \nonumber \\ 
B_{12} &=& \frac{1}{2} (\hat{E}_{0000}^{(+)} + 2\hat{E}_{0000}^{(0)} + \hat{E}_{0000}^{(-)} 
- \hat{E}_{1100}^{(+)} - 2\hat{E}_{1100}^{(0)} - \hat{E}_{1100}^{(-)} + \hat{E}_{0011}^{(+)} - \hat{E}_{0011}^{(-)} \nonumber \\
&& \quad + \hat{E}_{1111}^{(+)} - \hat{E}_{1111}^{(-)}) \nonumber \\ 
B_{13} &=& \frac{1}{2} (\hat{E}_{0000}^{(+)} + \hat{E}_{0000}^{(-)} + \hat{E}_{1100}^{(+)} + \hat{E}_{1100}^{(-)} 
+ \hat{E}_{0011}^{(+)} + \hat{E}_{0011}^{(-)} + \hat{E}_{1111}^{(+)} + \hat{E}_{1111}^{(-)}) + \, \hat{\upsilon} \nonumber \\
B_{14} &=& \frac{1}{2} (\hat{E}_{0000}^{(+)} + \hat{E}_{0000}^{(-)} - \hat{E}_{1100}^{(+)} - \hat{E}_{1100}^{(-)} 
+ \hat{E}_{0011}^{(+)} + \hat{E}_{0011}^{(-)} - \hat{E}_{1111}^{(+)} - \hat{E}_{1111}^{(-)}) \nonumber \\
B_{15} &=& \frac{1}{2} (\hat{E}_{0000}^{(+)} + \hat{E}_{0000}^{(-)} + \hat{E}_{1100}^{(+)} + \hat{E}_{1100}^{(-)} 
- \hat{E}_{0011}^{(+)} - \hat{E}_{0011}^{(-)} - \hat{E}_{1111}^{(+)} - \hat{E}_{1111}^{(-)}) \nonumber \\
B_{16} &=& \frac{1}{2} (\hat{E}_{0000}^{(+)} + \hat{E}_{0000}^{(-)} - \hat{E}_{1100}^{(+)} - \hat{E}_{1100}^{(-)} 
- \hat{E}_{0011}^{(+)} - \hat{E}_{0011}^{(-)} + \hat{E}_{1111}^{(+)} + \hat{E}_{1111}^{(-)}) \nonumber \\
B_{17} &=& \frac{1}{2} (\hat{E}_{0000}^{(+)} - \hat{E}_{0000}^{(-)} + \hat{E}_{1100}^{(+)} - \hat{E}_{1100}^{(-)} 
- \hat{E}_{0011}^{(+)} - 2\hat{E}_{0011}^{(0)} - \hat{E}_{0011}^{(-)} \nonumber \\  
&& \quad + \hat{E}_{1111}^{(+)} + 2\hat{E}_{1111}^{(0)} + \hat{E}_{1111}^{(-)}) \nonumber \\
B_{18} &=& \frac{1}{2} (\hat{E}_{0000}^{(+)} - \hat{E}_{0000}^{(-)} + \hat{E}_{1100}^{(+)} - \hat{E}_{1100}^{(-)} 
+ \hat{E}_{0011}^{(+)} + 2\hat{E}_{0011}^{(0)} + \hat{E}_{0011}^{(-)} \nonumber \\   
&& \quad - \hat{E}_{1111}^{(+)} - 2\hat{E}_{1111}^{(0)} - \hat{E}_{1111}^{(-)}) \nonumber \\
B_{19} &=& \frac{1}{2} (\hat{E}_{0000}^{(+)} - \hat{E}_{0000}^{(-)} - \hat{E}_{1100}^{(+)} + \hat{E}_{1100}^{(-)} 
+ \hat{E}_{0011}^{(+)} + 2\hat{E}_{0011}^{(0)} + \hat{E}_{0011}^{(-)}  \nonumber \\  
&& \quad + \hat{E}_{1111}^{(+)} + 2\hat{E}_{1111}^{(0)} + \hat{E}_{1111}^{(-)}) + \, \hat{\upsilon} \nonumber \\
B_{20} &=& \frac{1}{2} (\hat{E}_{0000}^{(+)} - \hat{E}_{0000}^{(-)} - \hat{E}_{1100}^{(+)} + \hat{E}_{1100}^{(-)} 
- \hat{E}_{0011}^{(+)} - 2\hat{E}_{0011}^{(0)} - \hat{E}_{0011}^{(-)} \nonumber \\   
&& \quad - \hat{E}_{1111}^{(+)} - 2\hat{E}_{1111}^{(0)} - \hat{E}_{1111}^{(-)}) - \, \hat{\upsilon} \; .
\eeq

\subsection{Identification of lattice vectors at the intersection point\label{ident}}
In \cite{NW99} it was shown that the conformal field theories associated with the K3 geometries
at ${\cal K} (\Z^4,0)$ and $\Z_4 (\frac{1}{\sqrt{2}} D_4, B^*)$ can be identified and therefore signify the same point
in K3 moduli space. A direct proof that the geometries associated with these theories are indeed the same
is given by the following identification of the
respective lattices of both theories which  also identifies the fourplanes placed in these lattices
\beqn
\begin{array}{ll}
A_{1} \cong B_{6}  \qquad \qquad \qquad  & A_{2} \cong -B_{5}   \\
A_{3} \cong -B_{4}   & A_{4} \cong -B_{15}  \\
A_{5} \cong B_{2}    & A_{6} \cong -B_{13}  \\
A_{7} \cong B_{14}   & A_{8} \cong B_{16}   \\
A_{9} \cong -B_{1}   & A_{10} \cong B_{3}   \\
A_{11} \cong B_{8}   & A_{12} \cong B_{7}   \\
A_{13} \cong -B_{9}  & A_{14} \cong -B_{10} \\
A_{15} \cong B_{11}  & A_{16} \cong B_{12}  \\
A_{17} \cong -B_{17} & A_{18} \cong -B_{18} \\
A_{19} \cong B_{19}  & A_{20} \cong -B_{20} \\
A_{21} \cong -B_{21} & A_{22} \cong B_{22}  \\
A_{23} \cong B_{23}  & A_{24} \cong B_{24}  \; . 
\end{array}
\label{identification}
\eeqn
As this is a highly symmetric point in moduli space this identification is certainly
only one of a great variety of possible ones.

\Proof In order to prove the above statement \mref{identification} 
we first need to express the generators of one
lattice, let's say the ones of the lattice $\Gamma({\cal K} (\Z^4,0))$ of ${\cal K} (\Z^4,0)$, 
in terms of the respective basis, the $A$'s
in this case. Using the above equivalence we can, hence, find the vectors equivalent to
these generators in the $\R$--span of the other lattice, $\mbox{span}_{\R} ( B_i; \, i =1, \dots , 24 )$. 
These vectors have to be shown to be lattice vectors
in $\Gamma(\Z_4 \frac{1}{\sqrt{2}} D_4, B^*)$ again. Now, as already explained in the beginning 
of section \ref{section2}, both lattices are known to be even selfdual 
of signature $(4,20)$ and are hence unique up to isometries. 
As both basis $A_i$ and $B_j$ consist of pairwise orthogonal vectors and the identification
\mref{identification} preserves the norm of these basis vectors, the identification preserves the scalar product
on the whole lattice. Thus, as this set of vectors 
is known to be generators for one of the two lattices, it has to generate the other as well.
We have performed the explicit calculation for the whole set of generators of $\Gamma({\cal K} (\Z^4,0))$
written down in \mref{Z2_eq1} and \mref{Z2_eq2}. A choice of typical examples of this calculation 
can be found in appendix \ref{appA}.

On the other hand the fourplanes are spanned by $A_{21}$, $A_{22}$, $A_{23}$, and $A_{24}$ in
$\Gamma({\cal K} (\Z^4,0))$ and $B_{21}$, $B_{22}$, $B_{23}$, and $B_{24}$ 
in $\Gamma(\Z_4 (\frac{1}{\sqrt{2}} D_4, B^*))$, 
and are thus identified.
Hence we have given a simultaneous isomorphism of both lattices and both fourplanes
which describe the two theories ${\cal K} (\Z^4,0)$ and $\Z_4 (\frac{1}{\sqrt{2}} D_4, B^*)$.
This proves that both these theories signify the same point in K3 moduli space. \hspace{0.2cm} $\Box$



\section{Identification of CFTs at $(\hat{2})^4$ Gepner point}

Let us now turn to the conformal field theoretic identification at the $(\hat{2})^4$ Gepner point.
\cite{NW99,Wend00} have shown that two orbifold CFTs, the $\Z_2$ orbifold ${\cal K} (\Z^4,0)$ 
and the $\Z_4$ orbifold $\Z_4 (\frac{1}{\sqrt{2}} D_4, B^*)$ (now regarded as CFTs), 
coincide with the same Gepner like theory
$(\hat{2})^4$, a certain $\Z_2$ orbifold of the Gepner model $(2)^4$.
The last identification has been achieved via first identifying the two theories
${\cal K} (\frac{1}{\sqrt{2}} D_4, B^*)$ and $(\tilde{2})^4$, a $(\Z_2)^2$ orbifold of $(2)^4$, 
and then showing that the same
$\Z_2$ orbifold action transforms both theories 
to the above two theories $\Z_4 (\frac{1}{\sqrt{2}} D_4, B^*)$
and $(\hat{2})^4$. 
We now want to give this identification explicitely.

First recall that certain superconformal field theories at $c=6$ can be identified
using the following steps if they have a specifically enlarged symmetry algebra (\cite{NW99,Wend00}):
\begin{itemize}
 \item The partition function has to agree in each sector.
 \item The current part of the holomorphic symmetry algebra ${\cal A}_h$, 
i.e.\ the algebra of the fields with
$(h,\hb)=(1,0)$ has to agree. For the argument to work, the $su(2)_1^2$ part of the symmetry algebra
which originates in the $N=4$ supersymmetry structure has to be enhanced by at least a $u(1)^4$.
Hence $u(1)^6 \subset {\cal A}_h$.
The same applies to the antiholomorphic symmetry algebra ${\cal A}_{\hb}$.
 \item Denote the $U(1)$ currents in $su(2)_1^2$ as $J^{(1)}$, $J^{(2)}$, the 
$U(1)$ currents in $u(1)^4$ as $j^1, \dots, j^4$. Define the ``bosonic'' Hilbert subspace
\beq
{\cal H}_b &:=& \left\{ |\varphi\rangle \in {\cal H} \left| 
J^{(k)}_0 |\varphi\rangle = 0 \quad \forall \; k \in \{1,2\} \right. \right\} \; .
\eeq
Then the charge lattice w.r.t.\ the subalgebra $u(1)^4$
\beq
\Gamma_b &:=& \left\{ \gamma \in \R^{d,d} \left| \; \exists |\varphi\rangle \in {\cal H}_b : \;
j^k_0 |\varphi\rangle = \gamma^k  |\varphi\rangle \quad \forall \; k \in \{1, \dots 4 \} \right. \right\}
\eeq
has to be isomorphic to the same selfdual lattice in both theories as well. 
It suffices to show this agreement for 
a set of generators $|\varphi\rangle$ of this selfdual lattice.
\end{itemize}
The main idea of the proof of this statement (\cite{NW99,Wend00}) is that
w.r.t.\ the $u(1)^4$ subalgebra the bosonic part of these theories can be viewed as a toroidal
theory in $d=4$ dimensions, which is uniquely determined by its charge lattice. 
The identification of the partition function
and the supersymmetry algebra then determine the complete supersymmetric theories to be isomorphic.

In the special case of our three theories the partition functions have already been shown to agree and been 
written down
in \cite{NW99,Wend00}. We now elaborate the symmetry algebra for the three theories which is
enhanced by $u(1)^4$ and the lattices of $(1/4,1/4)$ Ramond groundstate fields explicitely.
The $(1/4,1/4)$ Ramond groundstate fields generate ${\cal H}_b$ as they are the lowest components
of the supersymmetric groundstate fields.

\subsection{The symmetry algebra of ${\cal K} (\Z^4,0)$\label{symalg31}}
As we already know the supersymmetric generators in the symmetry algebra
we still need to extract the additional holomorphic vertex operators with 
$(h,\hb) = (1,0)$ in ${\cal K} (\Z^4,0)$.
The charge lattice of ${\cal K} (\Z^4,0)$ is given by
\beq
(P_r(\mu,\lambda), P_l(\mu,\lambda)) &=& \frac{1}{\sqrt{2}} 
\left(\mu + \lambda, \mu  -\lambda \right) = \left( p + \frac{1}{2} w,  p - \frac{1}{2} w \right) \; ,
\eeq
with $w = \sqrt{2} \lambda \in \sqrt{2} \Lambda = \sqrt{2} \Z^4$ 
and $p=\frac{1}{\sqrt{2}} \mu \in \frac{1}{\sqrt{2}} \Lambda^* = \frac{1}{\sqrt{2}} \Z^4$.

Hence we get four additional holomorphic $U(1)$ currents which are invariant under $\Z_2$ action
$e_i \mapsto -e_i \; i \in \{1,\dots,4\}$ (using the convention $V_P = V_{\lambda, \mu}$)
\beq
U_i &=& V^{\hmbox{inv}}_{e_i,e_i} = V^t_{e_i,e_i} +  V^t_{-e_i,-e_i} \quad \quad i = 1, \dots, 4 \; .
\eeq
The cocycle factors for a $u(1)^4$ algebra of holomorphic vertex operators are naturally 1.
Hence we get the OPE
\beq
U_i (z) U_j(w) = 2 \delta_{ij} (z-w)^{-2} + O(1) \; .
\eeq
The total, enhanced holomorphic symmetry algebra can be diagonalised as
(with the toroidal bosonic currents $j^{(k)}_{\pm} = \frac{1}{\sqrt{2}} (j^{2k-1} \pm i \, j^{2k})$
and their superpartners $\psi^{(k)}_{\pm} = \frac{1}{\sqrt{2}} (\psi^{2k-1} \pm i \, \psi^{2k})$)
\beq \renewcommand{\arraystretch}{1.5}
\begin{array}{lll}
J^{\prime} := (\psi_+^{(1)} \psi_-^{(1)}  + \psi_+^{(2)} \psi_-^{(2)}) \quad 
& J^{\prime +} := \sqrt{2} \psi_+^{(1)} \psi_+^{(2)} \quad &
J^{\prime -} := \sqrt{2} \psi_-^{(2)} \psi_-^{(1)} \\
A^{\prime} := (\psi_+^{(1)} \psi_-^{(1)}  - \psi_+^{(2)} \psi_-^{(2)}) 
& A^{\prime +} := \sqrt{2} \psi_+^{(1)} \psi_-^{(2)} &
A^{\prime -} := \sqrt{2} \psi_+^{(2)} \psi_-^{(1)} \\
\hline
P^{\prime} = \frac{1}{2} (U_1+U_2+U_3+U_4) \; &&\\
Q^{\prime} = \frac{1}{2} (U_1+U_2-U_3-U_4) \; && \\
R^{\prime} = \frac{1}{2} (U_1-U_2+U_3-U_4) \; && \\
S^{\prime} = \frac{1}{2} (U_1-U_2-U_3+U_4) \, . &&
\end{array} 
\eeq \renewcommand{\arraystretch}{1}

\noindent
This diagonalisation of the $U(1)$ currents has the advantage that $P^{\prime}$ and $Q^{\prime}$ are already
invariant under the $\Z_4$ operation
\beqn \label{Z4operation}
e_1 \mapsto e_2 \quad \quad e_2 \mapsto -e_1 \nonumber \\ 
e_3 \mapsto -e_4 \quad \quad e_4 \mapsto e_3 \; .
\eeqn
This $\Z_4$ operation acts the same way on the currents $j^i$ and their superpartners $\psi^i$.

\subsection{The symmetry algebra of $\Z_4 (\frac{1}{\sqrt{2}} D_4, B^*)$}
Remember the definition of $D_4$ and $B^*$ in \mref{Bfield}.
It is now easier to study the symmetry algebra of the $\Z_2$ orbifold ${\cal K} (\frac{1}{\sqrt{2}} D_4, B^*)$
first and to get the symmetry algebra of $\Z_4 (\frac{1}{\sqrt{2}} D_4, B^*)$
by an explicit orbifold action thereafter.

Hence, let us first look for additional holomorphic vertex operators in ${\cal K} (\frac{1}{\sqrt{2}} D_4, B^*)$.
Due to $\Lambda = \frac{1}{\sqrt{2}} D_4$ the left charges of these are given as roots out of the root system of $D_4$
\beq
(P_l, P_r) &=& \frac{1}{\sqrt{2}} 
\left(\mu - B^* \lambda + \lambda, \mu - B^* \lambda -\lambda \right) = (\alpha_i, 0) \qquad  \alpha_i \in D_4 \; ,
\eeq
which leads to the following cocycle factor on the root system of $D_4$ (\cite{GO86,Jun92})
\beq
c_{\alpha_2}(-\alpha_1) 
&=& \exp \left[ \frac{i}{2} \pi \alpha_1^t B^* \alpha_2 \right]  \; . 
\eeq
The roots of $D_4$ are given as linear combinations of the unit vectors $\alpha = e_i \pm e_j$. Hence, we can change 
the basis of the  vertex operators in the holomorphic algebra to the more useful linear
combination (following \cite{NW99})
\beq
W_{ij}^{\pm} := \frac{1}{2} \left(V^{\hmbox{inv}}_{e_i+e_j} \pm V^{\hmbox{inv}}_{e_i-e_j} \right) \; ,
\eeq
where 
\beq
V^{\hmbox{inv}}_{\alpha} (z) = (V^{\hmbox{torus}}_{\alpha} + V^{\hmbox{torus}}_{-\alpha} )
\eeq
is the $\Z_2$ invariant linear combination of the torus vertex operators $V^{\hmbox{torus}}$. 
We directly observe the symmetry
of the indices $W_{ij}^{\pm} = W_{ji}^{\pm}$.

One can now calculate the OPEs of all the $W_{ij}^{\pm}$ and rewrite these again, in order to make the
full $su(2)_1^6$ symmetry visible
\beqn \label{algebra2} \renewcommand{\arraystretch}{1.5}
\begin{array}{lll}
J := (\psi_+^{(1)} \psi_-^{(1)}  + \psi_+^{(2)} \psi_-^{(2)}) \quad & J^+ := \sqrt{2} \psi_+^{(1)} \psi_+^{(2)} \quad &
J^- := \sqrt{2} \psi_-^{(2)} \psi_-^{(1)} \\
A := (\psi_+^{(1)} \psi_-^{(1)}  - \psi_+^{(2)} \psi_-^{(2)}) & A^+ := \sqrt{2} \psi_+^{(1)} \psi_-^{(2)} &
A^- := \sqrt{2} \psi_+^{(2)} \psi_-^{(1)} \\
\hline
P := W^-_{14} + W^-_{23} & \multicolumn{2}{l}{ P^\pm := \frac{1}{\sqrt{2}} \left( ( W^+_{12} + W^+_{34}) \pm 
i  (W^-_{24} - W^-_{13}) \right) \;} \\
Q := W^+_{12} - W^+_{34} & \multicolumn{2}{l}{ Q^\pm := \frac{1}{\sqrt{2}} \left( ( W^-_{24} + W^-_{13}) \pm 
i  (W^-_{14} - W^-_{23}) \right) \;} \\
R := W^+_{14} + W^+_{23} & \multicolumn{2}{l}{ R^\pm := \frac{1}{\sqrt{2}} \left( ( W^+_{24} - W^+_{13}) \pm 
i  (W^-_{12} + W^-_{34}) \right) \;} \\
S := W^+_{24} + W^+_{13} & \multicolumn{2}{l}{ S^\pm := \frac{1}{\sqrt{2}} \left( ( W^-_{12} - W^-_{34}) \pm 
i  (W^+_{14} - W^+_{23}) \right) \, ,}
\end{array} 
\eeqn \renewcommand{\arraystretch}{1}

\noindent
where the $su(2)_1$ currents ${\cal J}$, ${\cal J}^{\pm}$ are normalised to fulfil the following OPEs
\beqn 
{\cal J}(z) \; {\cal J}^{\pm} (w) &\sim& \pm \, 2 \, (z-w)^{-1} \, {\cal J}^{\pm} (w) \nonumber \\
{\cal J}^+(z) \; {\cal J}^- (w) &\sim& \pm \, 2 \, (z-w)^{-2} + 2\,(z-w)^{-1} \, {\cal J} (w) \nonumber \\
{\cal J}(z) \; {\cal J} (w) &\sim& \pm \, 2 \, (z-w)^{-2} \; . \label{su2_norm}
\eeqn

Now, we turn to the symmetry algebra of the $\Z_4 (\frac{1}{\sqrt{2}} D_4, B^*)$ model.
\cite{NW99} have shown that it is enhanced to 
$su(2)^2 \otimes u(1)^4$ which is just given by the symmetry currents of
\mref{algebra2} which are invariant under the $\Z_4$ operation \mref{Z4operation}. These are
\beq
J, \; J^{\pm} ; \quad P, \; P^{\pm} ; \quad A; \, Q; \, R; \, S \; .
\eeq
There are no new holomorphic $(1,0)$ currents in the twisted sectors.

\subsection{Non--twisted groundstate $(\frac{1}{4},\frac{1}{4})$ fields}
The torus theory contains eight groundstate $(\frac{1}{4},\frac{1}{4})$ fields in the Ramond sector.
These generate the Clifford algebra of groundstates for the Ramond sector. They can be diagonalised
to have non-vanishing charges only w.r.t.\ to one of the currents $J$, $A$ (as in \mref{algebra2}) 
and its respective antiholomorphic counterpart
\begin{itemize}
\item $E_J^{\pm}$ having charges $(\pm 1, \pm 1)$ w.r.t.\ $J$, $\bar{J}$
\item $F_J^{\pm}$ having charges $(\pm 1, \mp 1)$ w.r.t.\ $J$, $\bar{J}$
\item $E_A^{\pm}$ having charges $(\pm 1, \pm 1)$ w.r.t.\ $A$, $\bar{A}$
\item $F_A^{\pm}$ having charges $(\pm 1, \mp 1)$ w.r.t.\ $A$, $\bar{A}$.
\end{itemize}
All of these fields survive the $\Z_2$ orbifolding, but only six of them,
\beq
E_J^{\pm}, \quad F_J^{\pm}, \quad E_A^{\pm} \; ,
\eeq
survive the above $\Z_4$ orbifolding with the action of equation \mref{Z4operation}.
As the two torus symmetry fields $J$ and $A$ are made up of free fermions (the superpartners of
the bosonic currents of the torus theory) a representation for these eight groundstate fields can
be written down in terms of a suitable bosonisation.

\subsection{The OPE of vertex operators with groundstate twistfields}
In \cite{Lep85,Jun92,NW01}
it was shown that the general OPE of a vertex operator of the original torus theory
with a groundstate twistfield of a $\Z_N$ orbifold theory looks like
\beq
V^{\hmbox{torus}}_{P(\mu,\lambda)} (z) \, T_f^l(w) &=& (z-w)^h (\zb-\wb)^{\hb} g (P_L, P_R) \,
\zeta_N^{\mu (Nx_f) l}  \,  T_{f^{\prime}}^l(w) + \dots
\eeq
with $\zeta_N = \exp{2\pi i /N}$ and the translated fixpoint
\beq
x_{f^{\prime}} &=& x_f + [(1-\theta)^{-1} \lambda] = x_f \left[\frac{1}{n(f)} 
\sum_1^{n(f)-1} k \theta^k \lambda \right] \; .
\eeq
$\theta$ is the matrix of the representation of the orbifold group generator on the torus lattice, e.g.\ equation
\mref{Z4operation} for the above $\Z_4$ orbifold. $f = (1-\theta) \; x_f \in \Lambda$ signifies the
fixpoint the twistfield lives at; $x_f\in I$ gives the location of that fixed point, where $I$ can be taken as
a subgroup of order $n(f)$ of $H_1(T, \R) / H_1(T, \Z)$.

The calculation in \cite{Jun92} makes use of the mode expansion of the fields. There the coupling parameter
$g (P_L, P_R)$, which is independent of the position of the fixpoint, 
is determined by a careful treatment of the zero modes to be
\beq
g (P_L, P_R) &=& e^{\pi i p^t (1-\theta)^{-1} w} \; g^{\prime}_l (P_L, P_R) 
\eeq
with the vertex operator coupling constant ($d/2$ signifies the complex dimension)
\beq
g^{\prime}_l (P_L, P_R) &=& \prod_{\mu = 1}^{d/2} \; \delta(l \, k_{\mu})^{-(h^{\mu} + \hb^{\mu})} \nonumber \\
\delta(k_{\mu}) &=& N^2 \prod_{a=1}^{N-1} \; \left(2 \sin \frac{\pi a}{N} \right)^{-2 \cos (2\pi a k_{\mu})} \nonumber \\
h^{\mu} &=& \frac{1}{2} ||P_L^{\mu}||^2, \qquad \hb^{\mu} = \frac{1}{2} ||P_R^{\mu}||^2 \; .
\eeq

\subsection{Twisted groundstate $(\frac{1}{4},\frac{1}{4})$ fields in $\Z_4 (\frac{1}{\sqrt{2}} D_4, B^*)$
\label{Z4_twistfields}}
First regard the action of the additional symmetry generators on the groundstate twistfields of
this theory. We have the lattice $\Lambda = \frac{1}{\sqrt{2}} D_4 = \frac{1}{\sqrt{2}} \Lambda_d$ where 
$\Lambda_d$ will be the lattice numbering our fixpoints, i.e.\ $f \in \Lambda_d$. 
The representation of the orbifold group generator
is given by
\beqn \label{theta}
\theta_4 = 
\left( \begin{array}{cccc}
0 & -1 & 0 & 0 \\
1 & 0 & 0 & 0 \\
0 & 0 & 0 & 1 \\
0 & 0 & -1 & 0
\end{array} \right) \qquad
\theta_2 = \theta_4^2 =
\left( \begin{array}{cccc}
-1 & 0 & 0 & 0 \\
0 & -1 & 0 & 0 \\
0 & 0 & -1 & 0 \\
0 & 0 & 0 & -1
\end{array} \right) \; .
\eeqn

\subsubsection{$\Z_2$ twistfields}
For $\Z_2$ twistfields we have a coupling of 
\beq
g^{(N=2)} (\alpha, 0) = \frac{1}{4} \;  e^{\frac{1}{2} \pi i (\alpha+B^* \alpha)^t (1-\theta_2)^{-1} \alpha}  =  \frac{i}{4}
\eeq
due to $||\alpha||^2 =2$ and the antisymmetry of $B^*$. Furthermore,
the translation of the fixpoint in the OPE of the holomorphic vertex operator and
the twistfield is given by 
\beq
x_f \mapsto x_f + (1-\theta_2)^{-1} \lambda  = x_f + \frac{1}{2} \frac{\alpha}{\sqrt{2}} \; .
\eeq
Hence we get an OPE of
\beq
V_{e_i+e_j} (z) \; T_f^{(N=2)} (w) &=& (V^t_{e_i+e_j} (z) + V^t_{-e_i-e_j} (z))\; T_f^{(N=2)} (w) \nonumber \\
&=& g^{(N=2)} (\alpha, 0) (z-w)^{-1} \nonumber \\
&&\left( e^{i\pi(\alpha+B^* \alpha)^t (\sqrt{2} x_f)} +
\quad e^{-i\pi(\alpha+B^* \alpha)^t (\sqrt{2} x_f)} \right)  
\; T_{f^{\prime}}^{(N=2)} (w) \nonumber \\
&=& \frac{i}{2}  e^{i\pi(\alpha+B^* \alpha)^t (\sqrt{2} x_f)} \; T_{[x_f + \frac{1}{2 \sqrt{2}} \alpha]}^{(N=2)} (w) \; .
\eeq
Now using this OPE, we can diagonalise the groundstate twistfields w.r.t.\ the action of the
enhanced symmetry generators
\beq
E_P^{\pm} &=& \sum_{\delta_1 = \delta_2 \atop \delta_3 = \delta_4} (-1)^{\delta_3} 
T_{\underline{\delta}}^{(N=2)}
\pm i  \sum_{\delta_1 \neq \delta_2 \atop \delta_3 \neq \delta_4} (-1)^{\delta_2 + \delta_3} 
T_{\underline{\delta}}^{(N=2)} \nonumber \\
F_P^{\pm} &=& \sum_{\delta_1 \neq \delta_2 \atop \delta_3 = \delta_4} T_{\underline{\delta}}^{(N=2)}
\pm i  \sum_{\delta_1 = \delta_2 \atop \delta_3 \neq \delta_4} (-1)^{\delta_1} 
T_{\underline{\delta}}^{(N=2)} \nonumber \\
E_Q^{\pm} &=& \sum_{\delta_1 = \delta_2 \atop \delta_3 = \delta_4} T_{\underline{\delta}}^{(N=2)}
\pm i  \sum_{\delta_1 \neq \delta_2 \atop \delta_3 = \delta_4} (-1)^{\delta_3} 
T_{\underline{\delta}}^{(N=2)} \nonumber \\
F_Q^{\pm} &=& \sum_{\delta_1 = \delta_2 \atop \delta_3 \neq \delta_4} (-1)^{\delta_1+\delta_3}
T_{\underline{\delta}}^{(N=2)}
\pm i  \sum_{\delta_1 \neq \delta_2 \atop \delta_3 \neq \delta_4} (-1)^{\delta_2} 
T_{\underline{\delta}}^{(N=2)} \nonumber \\
E_R^{\pm} &=& \sum_{\delta_1 = \delta_2 \atop \delta_3 = \delta_4} (-1)^{\delta_1} 
T_{\underline{\delta}}^{(N=2)}
\pm i  \sum_{\delta_1 \neq \delta_2 \atop \delta_3 \neq \delta_4} T_{\underline{\delta}}^{(N=2)} \nonumber \\
F_R^{\pm} &=& \sum_{\delta_1 \neq \delta_2 \atop \delta_3 = \delta_4} (-1)^{\delta_2 + \delta_3} 
T_{\underline{\delta}}^{(N=2)}
\pm i  \sum_{\delta_1 = \delta_2 \atop \delta_3 \neq \delta_4} (-1)^{\delta_3} 
T_{\underline{\delta}}^{(N=2)} \nonumber \\
E_S^{\pm} &=& \sum_{\delta_1 = \delta_2 \atop \delta_3 = \delta_4} (-1)^{\delta_1 + \delta_3} 
T_{\underline{\delta}}^{(N=2)}
\pm i  \sum_{\delta_1 = \delta_2 \atop \delta_3 \neq \delta_4} T_{\underline{\delta}}^{(N=2)} \nonumber \\
F_S^{\pm} &=& \sum_{\delta_1 \neq \delta_2 \atop \delta_3 = \delta_4} (-1)^{\delta_1} 
T_{\underline{\delta}}^{(N=2)}
\pm i  \sum_{\delta_1 \neq \delta_2 \atop \delta_3 \neq \delta_4} (-1)^{\delta_3} 
T_{\underline{\delta}}^{(N=2)} \; ,
\eeq
where the index indicates the $U(1)$ current ($P$, $Q$, $R$ or $S$) this field is charged with. The 
respective holomorphic and antiholomorphic charges are $(\pm 1, \pm 1)$ 
for fields $E^{\pm}_{\; \cdot}$ and $(\pm 1, \mp 1)$ for fields $F^{\pm}_{\; \cdot}$. 
The charges w.r.t.\ the respective other three
currents as well as $J$ and $A$ vanish. The following ten fields are invariant under the $\Z_4$ action \mref{Z4operation}
and are, hence, contained in $\Z_4 (\frac{1}{\sqrt{2}} D_4, B^*)$ as well
\beq
E^{\pm}_P, \; F^{\pm}_P; \quad E^{\pm}_Q;  \quad E^{\pm}_R;  \quad E^{\pm}_S \; .
\eeq

\subsubsection{$\Z_4$ twistfields}
In the case of $\Z_4$ twistfields the vertex operator coupling amounts to
\beq
g^{(N=4) \; \prime}_l (\alpha, 0) = \frac{1}{8} \qquad \forall \; l = 1,3 \; .
\eeq
The total coupling constant $g^{(N=4)} (\alpha, 0)$ still depends on the order of the twist
as we will see below.
The fixpoint of the twistfield is translated due to the action of the vertex operator
\beq
x_f \mapsto x_{f^{\prime}} = x_f + (1-\theta_4^l)^{-1} \lambda = x_f + \frac{1}{2} (1+\theta_4^l) \frac{\alpha}{\sqrt{2}} \; .
\eeq
Hence we get an OPE for $l=1$ of
\beq
V_{e_i+e_j} (z) \; T_f^{(N=4) \; l=1} (w) &=& (V^t_{e_i+e_j} (z) + V^t_{-e_i-e_j} (z))\; T_f^{(N=4) \; l=1} (w) \nonumber \\
&=& g^{(N=4) \; \prime} (\alpha, 0) \; e^{\frac{1}{2} \pi i (\alpha+B^* \alpha)^t \frac{1}{2} (1 + \theta_4) \; 
\alpha}(z-w)^{-1}  \nonumber \\
&& \left( e^{i\pi(\alpha+B^* \alpha)^t (\sqrt{2} x_f)} +
e^{-i\pi(\alpha+B^* \alpha)^t (\sqrt{2} x_f)} \right) \; T_{f^{\prime}}^{(N=4) \; l=1} (w) \nonumber \\
&=& \frac{1}{4}  e^{i\pi(\alpha+B^* \alpha)^t (\sqrt{2} x_f)} 
e^{\frac{i \pi}{2}  \alpha^t L_1 \alpha}
\; T_{[x_f + \frac{1}{2\sqrt{2}} (1+\theta_4) \alpha]}^{(N=4) \; l=1} (w) 
\eeq
as well as for $l=3$ of
\beq
V_{e_i+e_j} (z) \; T_f^{(N=4) \; l=3} (w) &=& (V^t_{e_i+e_j} (z) + V^t_{-e_i-e_j} (z))\; T_f^{(N=4) \; l=3} (w) \nonumber \\
&=& g^{(N=4) \; \prime} (\alpha, 0) \; e^{\frac{1}{2} \pi i (\alpha+B^* \alpha)^t \frac{1}{2} (1 + \theta_4) \alpha} \;
(z-w)^{-1}   \nonumber \\
&& \left( e^{3i\pi(\alpha+B^* \alpha)^t (\sqrt{2} x_f)} +
e^{-3i\pi(\alpha+B^* \alpha)^t (\sqrt{2} x_f)} \right) \; T_{f^{\prime}}^{(N=4) \; l=3} (w) \nonumber \\
&=& \frac{1}{4}  e^{3i\pi(\alpha+B^* \alpha)^t (\sqrt{2} x_f)} 
e^{\frac{i \pi}{2} \alpha^t L_3 \alpha}
\; T_{[x_f + \frac{1}{2\sqrt{2}} (1-\theta_4) \alpha]}^{(N=4) \; l=3} (w) 
\eeq
with 
\beq
L_1 = \left( \begin{array}{cccc}
0 & -1 & 0 & 0 \\
1 & 0 & 0 & 0 \\
0 & 0 & 1 & 0 \\
0 & 0 & 0 & 1
\end{array} \right) \qquad
L_3 = \left( \begin{array}{cccc}
1 & 0 & 0 & 0 \\
0 & 1 & 0 & 0 \\
0 & 0 & 0 & -1 \\
0 & 0 & 1 & 0
\end{array} \right) \; .
\eeq
Using this OPE one can diagonalise the twistfields according to
\beq \renewcommand{\arraystretch}{1.5}
\begin{array}{|l||c|c|c|c|}
& q_A & q_Q & q_R & q_S \\
\hline \hline
N_1 := (T_{0000}^1 + T_{1100}^1) + i (T_{1111}^1 - T_{0011}^1) & + \frac{1}{2} & + \frac{1}{2} & + \frac{1}{2} & - \frac{1}{2} \, \\
\hline
N_2 := (T_{0000}^1 + T_{1100}^1) - i (T_{1111}^1 - T_{0011}^1) & + \frac{1}{2} & + \frac{1}{2} & - \frac{1}{2} & + \frac{1}{2} \, \\
\hline
N_3 := (T_{0000}^1 - T_{1100}^1) + i (T_{1111}^1 + T_{0011}^1) & + \frac{1}{2} & - \frac{1}{2} & + \frac{1}{2} & + \frac{1}{2} \, \\
\hline
N_4 := (T_{0000}^1 - T_{1100}^1) - i (T_{1111}^1 + T_{0011}^1) & + \frac{1}{2} & - \frac{1}{2} & - \frac{1}{2} & - \frac{1}{2} \, \\
\hline
N_5 := (T_{0000}^3 + T_{1100}^3) + i (T_{1111}^3 - T_{0011}^3) & - \frac{1}{2} & - \frac{1}{2} & + \frac{1}{2} & - \frac{1}{2} \, \\
\hline
N_6 := (T_{0000}^3 + T_{1100}^3) - i (T_{1111}^3 - T_{0011}^3) & - \frac{1}{2} & - \frac{1}{2} & - \frac{1}{2} & + \frac{1}{2} \, \\
\hline
N_7 := (T_{0000}^3 - T_{1100}^3) + i (T_{1111}^3 + T_{0011}^3) & - \frac{1}{2} & + \frac{1}{2} & + \frac{1}{2} & + \frac{1}{2} \, \\
\hline
N_8 := (T_{0000}^3 - T_{1100}^3) - i (T_{1111}^3 + T_{0011}^3) & - \frac{1}{2} & + \frac{1}{2} & - \frac{1}{2} & - \frac{1}{2} \, \\
\end{array} 
\eeq \renewcommand{\arraystretch}{1}

\noindent
where all of these are not charged under $J$ and $P$ and the antiholomorphic 
charges are just the same as the holomorphic ones.
The charge w.r.t.\ $A$ cannot be derived from the above OPE, 
as it is a current made up of the $(\frac{1}{2}, 0)$ fermions
of the torus theory. In appendix \ref{Rsector} we show that
a very careful calculation of the R--Sector partition function --- keeping the factors originating from
the bosonic and fermionic characters well apart --- reveals that the $\Z_4$ groundstate twistfields are made
up of a bosonic twistfield of conformal weight $h_b = \frac{3}{16}$ and a fermionic twistfield of 
conformal weight $h_f = \frac{1}{16}$. This perfectly well corresponds to the general formulae for bosonic
twistfields $h_b = \frac{1}{2} \frac{k}{N} (1-  \frac{k}{N})$ and the one for
fermionic twistfields $h_f = \frac{1}{2} \frac{k}{N}$ derived in \cite{Dix87} by
CFT methods. But one also observes, inspecting the
partition function, that these fields are uncharged w.r.t.\ the other current $J$ 
made up of $(\frac{1}{2}, 0)$ fermions. Hence they have to be charged under $A$. To see this 
we complexify the torus fields as in \ref{symalg31} and translate our specific orbifold action \mref{theta} 
to the complexified torus.
Then the orbifold action acts on both complex dimensions
of the torus in just the opposite way, i.e.\ with multiplication with phases which are complex
conjugate to each other. (This corresponds to the fact that the first two real dimensions
are rotated just the opposite way as the last two by \mref{theta}.)
However, following the derivation in \cite{Dix87}, that means that in the above formula for
the twisted fermionic conformal weights $h_f$ we have to take
$k$ for the first dimension and $-k$ for the second. But $J$ and $A$, as defined in \mref{algebra2}, 
measure the fermionic
content in both complex dimensions independently (with currents of the form $\psi_+^{(i)} \psi_-^{(i)}$
in the complex dimension $i$). $J$ adds the fermionic content, $A$ subtracts it. 
Knowing that (by convention) $l$ in $T^l$ refers to the first
complex dimension the claimed charges follow.

\subsection{Twisted groundstate $(\frac{1}{4},\frac{1}{4})$ fields in ${\cal K} (\Z_4, 0)$}
This time we only have $\Z_2$ twistfields with generator $\theta_2$ \mref{theta}. The lattice 
vectors are given by $e_i$, and hence
the fixpoint of the twistfield is translated due to the action of the vertex operator
\beq
x_f \mapsto x_{f^{\prime}} = x_f + (1-\theta_2)^{-1} e_i = x_f + \frac{1}{2} e_i \; .
\eeq
The fixpoints are elements of the finite group $x_f \in \frac{1}{2} \Z^4 / \Z^4$. 
The coupling is given as above
\beq
g^{(N=2)} (\alpha, 0) &=& \frac{1}{4} \; e^{\frac{1}{2} \pi i (\sqrt{2} e_i)^t \frac{1}{2} (\sqrt{2} e_i)} = \frac{i}{4} \; .
\eeq
It follows an OPE
(with $\Sigma_f = T_f^{(N=2)}$)
\beq
U_i (z) \; \Sigma_f (w) &=& (V^t_{e_i,e_i} (z) + V^t_{-e_i,-e_i} (z))\; \Sigma_f (w) \nonumber \\
&=& g^{(N=2)}(\alpha, 0) \; (\sqrt{2} e_i, 0) (z-w)^{-1}  \nonumber \\
&& \quad \left( e^{2\pi i(\frac{1}{\sqrt{2}} e_i)^t (\sqrt{2} x_f)} +
e^{2\pi i(-\frac{1}{\sqrt{2}} e_i)^t (\sqrt{2} x_f)} \right) 
\; \Sigma_{f^{\prime}} (w) \nonumber \\
&=& \frac{i}{2}  e^{i\pi e_i^t x_f} \; \Sigma_{[x_f + \frac{1}{2} e_i]} (w) \; .
\eeq

\noindent
This OPE yields the following diagonalisation of the twistfields ($N^{\prime}_{i \, / \, j}$ means:
$N^{\prime}_i$ resp. $N^{\prime}_j$)
\beq \renewcommand{\arraystretch}{2}
\begin{array}{|l||c|c|c|c|}
& q_P & q_Q & q_R & q_S \\
\hline \hline
{E^{\prime \, \pm}_P := (\Sigma_{0000}-\Sigma_{1100}-\Sigma_{0011}+\Sigma_{1111}-\Sigma_{1010}-\Sigma_{1001}-\Sigma_{0110}-\Sigma_{0101})
\atop \pm i (\Sigma_{1000}+\Sigma_{0100}+\Sigma_{0010}+\Sigma_{0001}-\Sigma_{1110}-\Sigma_{1101}-\Sigma_{1011}-\Sigma_{0111})} &
\pm 1 & 0 & 0 & 0 \, \\ \hline
{E^{\prime \, \pm}_Q := (\Sigma_{0000}-\Sigma_{1100}-\Sigma_{0011}+\Sigma_{1111}+\Sigma_{1010}+\Sigma_{1001}+\Sigma_{0110}+\Sigma_{0101})
\atop \pm i (\Sigma_{1000}+\Sigma_{0100}-\Sigma_{0010}-\Sigma_{0001}+\Sigma_{1110}+\Sigma_{1101}-\Sigma_{1011}-\Sigma_{0111})} &
0 & \pm 1 & 0 & 0 \, \\ \hline
{E^{\prime \, \pm}_R := (\Sigma_{0000}+\Sigma_{1100}+\Sigma_{0011}+\Sigma_{1111}-\Sigma_{1010}+\Sigma_{1001}+\Sigma_{0110}-\Sigma_{0101})
\atop \pm i (\Sigma_{1000}-\Sigma_{0100}+\Sigma_{0010}-\Sigma_{0001}+\Sigma_{1110}-\Sigma_{1101}+\Sigma_{1011}-\Sigma_{0111})} &
0 & 0 & \pm 1 & 0 \, \\ \hline
{E^{\prime \, \pm}_S := (\Sigma_{0000}+\Sigma_{1100}+\Sigma_{0011}+\Sigma_{1111}+\Sigma_{1010}-\Sigma_{1001}-\Sigma_{0110}+\Sigma_{0101})
\atop \pm i (\Sigma_{1000}-\Sigma_{0100}-\Sigma_{0010}+\Sigma_{0001}-\Sigma_{1110}+\Sigma_{1101}+\Sigma_{1011}-\Sigma_{0111})} &
0 & 0 & 0 & \pm 1 \, \\ \hline
{N^{\prime}_{1 \, / \, 6} := (\Sigma_{0000}-\Sigma_{1100}+\Sigma_{0011}-\Sigma_{1111}-\Sigma_{1010}+\Sigma_{1001}-\Sigma_{0110}+\Sigma_{0101})
\atop \pm i (\Sigma_{1000}+\Sigma_{0100}+\Sigma_{0010}-\Sigma_{0001}-\Sigma_{1110}+\Sigma_{1101}+\Sigma_{1011}+\Sigma_{0111})} &
\pm \frac{1}{2} & \pm \frac{1}{2} & \pm \frac{1}{2} & \mp \frac{1}{2} \, \\ \hline
{N^{\prime}_{3 \, / \, 8} := (\Sigma_{0000}+\Sigma_{1100}-\Sigma_{0011}-\Sigma_{1111}-\Sigma_{1010}-\Sigma_{1001}+\Sigma_{0110}+\Sigma_{0101})
\atop \pm i (\Sigma_{1000}-\Sigma_{0100}+\Sigma_{0010}+\Sigma_{0001}+\Sigma_{1110}+\Sigma_{1101}-\Sigma_{1011}+\Sigma_{0111})} &
\pm \frac{1}{2} & \mp \frac{1}{2} & \pm \frac{1}{2} & \pm \frac{1}{2} \, \\ \hline
{N^{\prime}_{2 \, / \, 5} := (\Sigma_{0000}-\Sigma_{1100}+\Sigma_{0011}-\Sigma_{1111}+\Sigma_{1010}-\Sigma_{1001}+\Sigma_{0110}-\Sigma_{0101})
\atop \pm i (\Sigma_{1000}+\Sigma_{0100}-\Sigma_{0010}+\Sigma_{0001}+\Sigma_{1110}-\Sigma_{1101}+\Sigma_{1011}+\Sigma_{0111})} &
\pm \frac{1}{2} & \pm \frac{1}{2} & \mp \frac{1}{2} & \pm \frac{1}{2} \, \\ \hline
{N^{\prime}_{4 \, / \, 7} := (\Sigma_{0000}+\Sigma_{1100}-\Sigma_{0011}-\Sigma_{1111}+\Sigma_{1010}+\Sigma_{1001}-\Sigma_{0110}-\Sigma_{0101})
\atop \pm i (-\Sigma_{1000}+\Sigma_{0100}+\Sigma_{0010}+\Sigma_{0001}+\Sigma_{1110}+\Sigma_{1101}+\Sigma_{1011}-\Sigma_{0111})} &
\pm \frac{1}{2} & \mp \frac{1}{2} & \mp \frac{1}{2} & \mp \frac{1}{2} \, \\
\end{array} 
\eeq \renewcommand{\arraystretch}{1}

\noindent
where all of these are not charged under $J$ and $A$ and the antiholomorphic charges are just the same as the holomorphic ones.

\subsection{Fields of the Gepner model $(\hat{2})^4$}
It is easier to describe the field content of the more symmetric Gepner model  $(\tilde{2})^4$
with $\hat{su}(2)_1^6$ symmetry algebra first and then to derive the field content of
$(\hat{2})^4$ by orbifolding. Details about the calculations in Gepner models
and especially about how to compute the $(2)$ superminimal model as a tensor product
of a $c=1$ theory and an Ising model can be found in \cite{Wend00}.
Following \cite{NW99}, we define $X_{ij}$ to be the Gepner model 
field with $\Phi^0_{4,2;0,0}$ as the $i$th and $j$th tensor
factors and $\Phi^0_{0,0;0,0}$ elsewhere, and $Y_{ij}$ to be the Gepner Model field 
with $\Phi^0_{-2,2;0,0}$ as the $i$th and $j$th tensor
factors and $\Phi^0_{2,2;0,0}$ elsewhere. Furthermore let $J_i$ be the $U(1)$ current of the $i$th minimal model.
Then the complete $\hat{su}(2)_1^6$ symmetry algebra of $(\tilde{2})^4$ is given by
\beqn \label{algebra3} \renewcommand{\arraystretch}{1.5}
\begin{array}{lll}
J^{\prime \prime} := J_1 + J_2 + J_3 + J_4 \quad & 
\multicolumn{2}{l}{J^{\prime \prime \pm} := \sqrt{2} \; (\Phi^0_{\mp 2, 2; 0,0})^{\otimes 4} \;} \\
A^{\prime \prime} := J_1 + J_2 - J_3 - J_4 & 
A^{\prime \prime +} := \sqrt{2} \; Y_{12} \quad &  A^{\prime \prime -} := \sqrt{2} \; Y_{34} \; \\
\hline
P^{\prime \prime} := J_1 - J_2 + J_3 - J_4 & P^{\prime \prime +} := \sqrt{2} \; Y_{13} &  P^{\prime \prime -} := \sqrt{2} \; Y_{24} \; \\
Q^{\prime \prime} := J_1 - J_2 - J_3 + J_4 & Q^{\prime \prime +} := \sqrt{2} \; Y_{14} &  Q^{\prime \prime -} := \sqrt{2} \; Y_{23} \; \\
R^{\prime \prime} := i(X_{13} - X_{24}) & \multicolumn{2}{l}{R^{\prime \prime \pm} := 
\frac{i}{\sqrt{2}} (X_{14} + X_{23}) \pm \frac{1}{\sqrt{2}} (X_{12} + X_{34}) \;} \\
S^{\prime \prime} := i(X_{13} + X_{24}) & \multicolumn{2}{l}{S^{\prime \prime \pm} := 
\frac{i}{\sqrt{2}} (X_{14} - X_{23}) \mp \frac{1}{\sqrt{2}} (X_{12} - X_{34}) \; ,}
\end{array} 
\eeqn \renewcommand{\arraystretch}{1}
normalised as \mref{su2_norm}.

The $(\frac{1}{4},\frac{1}{4})$ fields corresponding to Ramond groundstates can thus be diagonalised 
w.r.t.\ the action of the above currents
\beq
E^{\prime \prime \, \pm}_J &=& (\Phi^0_{\mp 1,\mp 1;\mp 1,\mp 1})^{\otimes 4} \nonumber \\
F^{\prime \prime \, \pm}_J &=& (\Phi^0_{\mp 1,\mp 1;\pm 1,\pm 1})^{\otimes 4} \nonumber \\
E^{\prime \prime \, \pm}_A &=& (\Phi^0_{\mp 1,\mp 1;\mp 1,\mp 1})^{\otimes 2} \otimes 
(\Phi^0_{\pm 1,\pm 1;\pm 1,\pm 1})^{\otimes 2} \nonumber \\
F^{\prime \prime \, \pm}_A &=& (\Phi^0_{\mp 1,\mp 1;\pm 1,\pm 1})^{\otimes 2} \otimes 
(\Phi^0_{\pm 1,\pm 1;\mp 1,\mp 1})^{\otimes 2} \nonumber \\
E^{\prime \prime \, \pm}_P &=& \Phi^0_{\mp 1,\mp 1;\mp 1,\mp 1} \otimes  \Phi^0_{\pm 1,\pm 1;\pm 1,\pm 1} \otimes 
\Phi^0_{\mp 1,\mp 1;\mp 1,\mp 1} \otimes  \Phi^0_{\pm 1,\pm 1;\pm 1,\pm 1} \nonumber \\
F^{\prime \prime \, \pm}_P &=& \Phi^0_{\mp 1,\mp 1;\pm 1,\pm 1} \otimes  \Phi^0_{\pm 1,\pm 1;\mp 1,\mp 1} \otimes 
\Phi^0_{\mp 1,\mp 1;\pm 1,\pm 1} \otimes  \Phi^0_{\pm 1,\pm 1;\mp 1,\mp 1} \nonumber \\
E^{\prime \prime \, \pm}_Q &=& \Phi^0_{\mp 1,\mp 1;\mp 1,\mp 1} \otimes  \Phi^0_{\pm 1,\pm 1;\pm 1,\pm 1} \otimes 
\Phi^0_{\pm 1,\pm 1;\pm 1,\pm 1} \otimes  \Phi^0_{\mp 1,\mp 1;\mp 1,\mp 1} \nonumber \\
F^{\prime \prime \, \pm}_Q &=& \Phi^0_{\mp 1,\mp 1;\pm 1,\pm 1} \otimes  \Phi^0_{\pm 1,\pm 1;\mp 1,\mp 1} \otimes 
\Phi^0_{\pm 1,\pm 1;\mp 1,\mp 1} \otimes  \Phi^0_{\mp 1,\mp 1;\pm 1,\pm 1} \nonumber \\
E^{\prime \prime \, \pm}_R &=& (\Phi^1_{2,1;2,1})^{\otimes 4} + (\Phi^1_{2,1;-2,-1})^{\otimes 4} \nonumber \\
&& \pm \left[(\Phi^1_{2,1;2,1} \otimes \Phi^1_{2,1;-2,-1})^{\otimes 2} 
- (\Phi^1_{2,1;-2,-1} \otimes \Phi^1_{2,1;2,1})^{\otimes 2} \right] \nonumber \\
F^{\prime \prime \, \pm}_R &=& (\Phi^1_{2,1;2,1})^{\otimes 2} \otimes (\Phi^1_{2,1;-2,-1})^{\otimes 2} 
+ (\Phi^1_{2,1;-2,-1})^{\otimes 2} \otimes (\Phi^1_{2,1;2,1})^{\otimes 2} \nonumber \\
&& \pm i \left[\Phi^1_{2,1;-2,-1} \otimes \Phi^1_{2,1;2,1} \otimes \Phi^1_{2,1;2,1} \otimes \Phi^1_{2,1;-2,-1} \right. \nonumber \\
&& \quad \left. + \Phi^1_{2,1;2,1} \otimes \Phi^1_{2,1;-2,-1} \otimes \Phi^1_{2,1;-2,-1} \otimes \Phi^1_{2,1;2,1} \right] \nonumber \\
E^{\prime \prime \, \pm}_S &=& (\Phi^1_{2,1;2,1})^{\otimes 4} - (\Phi^1_{2,1;-2,-1})^{\otimes 4} \nonumber \\
&& \mp \left[(\Phi^1_{2,1;2,1} \otimes \Phi^1_{2,1;-2,-1})^{\otimes 2} 
+ (\Phi^1_{2,1;-2,-1} \otimes \Phi^1_{2,1;2,1})^{\otimes 2} \right] \nonumber \\
F^{\prime \prime \, \pm}_S &=& (\Phi^1_{2,1;2,1})^{\otimes 2} \otimes (\Phi^1_{2,1;-2,-1})^{\otimes 2} 
- (\Phi^1_{2,1;-2,-1})^{\otimes 2} \otimes (\Phi^1_{2,1;2,1})^{\otimes 2} \nonumber \\
&& \pm i \left[\Phi^1_{2,1;-2,-1} \otimes \Phi^1_{2,1;2,1} \otimes \Phi^1_{2,1;2,1} \otimes \Phi^1_{2,1;-2,-1} \right. \nonumber \\
&& \quad \left. - \Phi^1_{2,1;2,1} \otimes \Phi^1_{2,1;-2,-1} \otimes \Phi^1_{2,1;-2,-1} \otimes \Phi^1_{2,1;2,1} \right] \; .
\eeq
As in section \ref{Z4_twistfields} the index indicates the $U(1)$ current this field is charged with, holomorphic resp.
antiholomorphic charges $(\pm 1, \pm 1)$ for $E^{\pm}_{\; \cdot}$ and $(\pm 1, \mp 1)$ for $F^{\pm}_{\; \cdot}$.

Now, as $(\hat{2})^4$ is gained from $(\tilde{2})^4$ by an $\Z_2$ orbifold generated by
\beq
[2^{\prime},2^{\prime},0,0]: \quad \bigotimes_{i=1}^4 \Phi^{l_i}_{m_i,s_i;\mb_i,\sbar_i} \; \mapsto \; 
e^{\frac{2\pi i}{8} (\mb_1 -m_1 -\mb_3+m_3)} \bigotimes_{i=1}^4 \Phi^{l_i}_{m_i,s_i;\mb_i,\sbar_i} \, ,
\eeq
the surviving $\hat{su}(2)_1^2 \otimes \hat{u}^4$ symmetry algebra of $(\hat{2})^4$ is given
by the currents
\beq
J, \; J^{\pm}; \quad A; \quad P, \; P^{\pm}; \quad Q; \quad R; \quad S.
\eeq
Of the above Ramond groundstate fields the following are invariant under the orbifold group action
\beq
E_J^{\prime \prime \,\pm}, \; F_J^{\prime \prime \,\pm}; \quad E_A^{\prime \prime \,\pm}; \quad E_P^{\prime \prime \,\pm}, \; 
F_P^{\prime \prime \,\pm}; \quad E_Q^{\prime \prime \,\pm}; \quad E_R^{\prime \prime \,\pm}; \quad E_S^{\prime \prime \,\pm}.
\eeq

The list of $(\frac{1}{4},\frac{1}{4})$ fields in $(\hat{2})^4$ has to be completed by the
following twistfields (w.r.t.\ the above orbifold construction)
\beq
T_1 &=& \Phi^1_{2,1;2,1} \otimes \Phi^0_{-1,-1;-1,-1} \otimes \Phi^1_{2,1;2,1} \otimes \Phi^0_{1,1;1,1} \nonumber \\
T_2 &=& \Phi^1_{2,1;2,1} \otimes \Phi^0_{1,1;1,1} \otimes \Phi^1_{2,1;2,1} \otimes \Phi^0_{-1,-1;-1,-1} \nonumber \\
T_3 &=& \Phi^0_{-1,-1;-1,-1} \otimes \Phi^1_{2,1;2,1} \otimes \Phi^0_{1,1;1,1} \otimes \Phi^1_{2,1;2,1} \nonumber \\
T_4 &=& \Phi^0_{1,1;1,1} \otimes \Phi^1_{2,1;2,1} \otimes \Phi^0_{-1,-1;-1,-1} \otimes \Phi^1_{2,1;2,1} \nonumber \\
T_5 &=& \Phi^1_{2,1;-2,-1} \otimes \Phi^0_{-1,-1;-1,-1} \otimes \Phi^1_{2,1;-2,-1} \otimes \Phi^0_{1,1;1,1} \nonumber \\
T_6 &=& \Phi^1_{2,1;-2,-1} \otimes \Phi^0_{1,1;1,1} \otimes \Phi^1_{2,1;-2,-1} \otimes \Phi^0_{-1,-1;-1,-1} \nonumber \\
T_7 &=& \Phi^0_{-1,-1;-1,-1} \otimes \Phi^1_{2,1;-2,-1} \otimes \Phi^0_{1,1;1,1} \otimes \Phi^1_{2,1;-2,-1} \nonumber \\
T_8 &=& \Phi^0_{1,1;1,1} \otimes \Phi^1_{2,1;-2,-1} \otimes \Phi^0_{-1,-1;-1,-1} \otimes \Phi^1_{2,1;-2,-1} \; .
\eeq
These can be diagonalised w.r.t.\ to the action of the invariant $U(1)$ currents
\beq \renewcommand{\arraystretch}{1.5}
\begin{array}{l||c|c|c|c|c|c}
& q_J & q_A & q_P & q_Q & q_R & q_S \\
\hline \hline
N^{\prime \prime}_{1 \, / \, 2} = T_3 \pm T_7 & 0 & +\frac{1}{2} & 0 & +\frac{1}{2} & \pm \frac{1}{2} & \pm \frac{1}{2} \\
\hline
N^{\prime \prime}_{3 \, / \, 4} = T_1 \mp T_5 & 0 & +\frac{1}{2} & 0 & -\frac{1}{2} & \pm \frac{1}{2} & \pm \frac{1}{2} \\
\hline
N^{\prime \prime}_{5 \, / \, 6} = T_4 \pm T_8 & 0 & -\frac{1}{2} & 0 & -\frac{1}{2} & \pm \frac{1}{2} & \pm \frac{1}{2} \\
\hline
N^{\prime \prime}_{7 \, / \, 8} = T_2 \mp T_6 & 0 & -\frac{1}{2} & 0 & +\frac{1}{2} & \pm \frac{1}{2} & \pm \frac{1}{2} 
\end{array}
\eeq \renewcommand{\arraystretch}{1}

\subsection{Explicit identification of the three theories}
Now we can perform the identification of the three CFTs $(\hat{2})^4$, ${\cal K} (\Z^4,0)$ and 
$\Z_4 (\frac{1}{\sqrt{2}} D_4, B^*)$, already proven in \cite{NW99}, explicitly
for the symmetry algebra and the lattice of $(\frac{1}{4},\frac{1}{4})$ fields. 
The identification of the symmetry algebra is not totally fixed (this reflects the high
amount of symmetry of these theories); one possible way of identifying the
currents is
\beq
J &\simeq& J^{\prime \prime} \;\simeq\; J^{\prime} \qquad J^{\pm} \;\simeq\; J^{\prime \prime \pm} \;\simeq\; J^{\prime \pm} \nonumber \\
A &\simeq& A^{\prime \prime} \;\simeq\; P^{\prime} \nonumber \\
P &\simeq& P^{\prime \prime} \;\simeq\; A^{\prime} \qquad P^{\pm} \;\simeq\; P^{\prime \prime \pm} \;\simeq\; A^{\prime \pm} \nonumber \\
Q &\simeq& Q^{\prime \prime} \;\simeq\; Q^{\prime} \nonumber \\
R &\simeq& R^{\prime \prime} \;\simeq\; R^{\prime} \nonumber \\
S &\simeq& S^{\prime \prime} \;\simeq\; S^{\prime} \; .
\eeq
Comparing the charges w.r.t.\ the above currents this leads to the following identification
of $(\frac{1}{4},\frac{1}{4})$ fields
\beqn \label{identi}
E_J^{\pm} &\simeq& E_J^{\prime \prime \pm} \;\simeq\; E_J^{\prime \pm} \qquad 
F_J^{\pm} \simeq F_J^{\prime \prime \pm} \;\simeq\; F_J^{\prime \pm} \nonumber \\
E_A^{\pm} &\simeq& E_A^{\prime \prime \pm} \;\simeq\; E_P^{\prime \pm} \nonumber \\
E_P^{\pm} &\simeq& E_P^{\prime \prime \pm} \;\simeq\; E_A^{\prime \pm} \qquad 
F_P^{\pm} \simeq F_P^{\prime \prime \pm} \;\simeq\; F_A^{\prime \pm} \nonumber \\
E_Q^{\pm} &\simeq& E_Q^{\prime \prime \pm} \;\simeq\; E_Q^{\prime \pm} \nonumber \\
E_R^{\pm} &\simeq& E_R^{\prime \prime \pm} \;\simeq\; E_R^{\prime \pm} \nonumber \\
E_S^{\pm} &\simeq& E_S^{\prime \prime \pm} \;\simeq\; E_S^{\prime \pm} \nonumber \\
N_i &=& N^{\prime}_i = N^{\prime \prime}_i \qquad \forall \; i \in \{1, \dots ,8 \} \; .
\eeqn
This identification leads to the important observation that the $\Z_2$ and $\Z_4$ 
orbifold subvarieties in K3 moduli space which intersect in one point 
as described earlier are orthogonal to each other. In order to see this it is important
to recall some facts about conformal deformation theory \cite{Kad78,Dix87,Wend00}. The possible exact
marginal deformation fields in these $c=6$ theories are not charged
under the $U(1)$ current $J$ of the SUSY algebra and can be generated as the $(1,1)$ superpartners
of $(\frac{1}{2},\frac{1}{2})$ NS fields which again can be generated via spectral flow
from $(\frac{1}{4},\frac{1}{4})$ Ramond fields. Now we have two types of deformation fields
within our orbifold theories. Deformation fields which originate from the original torus theory
generate all the deformations along the orbifold subvarieties and are well understood. The 
corresponding $(\frac{1}{4},\frac{1}{4})$ Ramond fields in ${\cal K} (\Z^4,0)$ are 
$E_A^{\prime \pm}$ and  $F_A^{\prime \pm}$, the ones in $\Z_4 (\frac{1}{\sqrt{2}} D_4, B^*)$
are $E_A^{\pm}$. On the other hand, we do not understand very much about 
deformations with twistfields. The $(\frac{1}{4},\frac{1}{4})$ Ramond fields corresponding
to these deformations are given by $E_P^{\prime \pm}$, $E_Q^{\prime \pm}$,
$E_R^{\prime \pm}$, $E_S^{\prime \pm}$, $N^{\prime}_i$
in ${\cal K} (\Z^4,0)$, and $E_P^{\pm}$, $F_P^{\pm}$, $E_Q^{\pm}$, $E_R^{\pm}$, $E_S^{\pm}$,
$N_i$ in $\Z_4 (\frac{1}{\sqrt{2}} D_4, B^*)$.
Now the orbifold group selection rules imply that the deformation fields originating from
the torus theory and the twisted deformation fields are orthogonal w.r.t.\ the
Zamolodchikov metric as any two point function of the two types has to vanish.
Now the identification in \mref{identi} implies that torus deformations of one 
theory are identified with twistfield deformations of the respective other.
This, the fact that deformations along the orbifold subvarieties are
generated by torus deformations and the orthogonality of the two types of deformation fields 
prove the above stated orthogonality of the $\Z_2$ and $\Z_4$ 
orbifold subvarieties.



\section{Conclusion}
In this paper we have given an explicit identification of the coordinates of 
the $\Z_2$ and $\Z_4$ orbifolds at the intersection of the $\Z_2$ and $\Z_4$ orbifold
subvarieties in the K3 moduli space. This is an important step
towards the exploration of the unknown parts of the K3 moduli space via
conformal deformation theory. It enables us to relate the coordinates
of the starting and end points of geodesics running between
the $\Z_2$ and $\Z_4$ orbifold subvarieties which we suppose to be the
most promising lines for a successful finite deformation.
Open questions in this direction are the explicit relation between the
geometric deformations and the deformation fields of the
respective conformal field theories as well as the
calculation of the deformed conformal weights and structure constants
of the CFTs themselves. Some results in this direction are presented
in \cite{Ebe01}; both questions are challenging, though, and
we are currently pursuing them further.

Furthermore, elaborating the explicit identification
of the respective CFTs at this point, especially on the level of the symmetry algebra
and the groundstate lattice, we have shown the important fact
that the two subvarieties are orthogonal to each other. 
As we can now express deformation twistfields of one orbifold theory by torus deformation
fields in another at this point of intersection, we can use this to probe the
deformation by twistfields. But twistfield deformations are exactly these
deformation fields we need to explore the unknown parts of the K3 moduli space, 
as described above. Hence, we may hope to find some clues at this intersection point
for the above mentioned unsolved problem how to identify the geometric deformations 
and the conformal deformation fields.
Additionally, this identification might also 
help to clarify the peculiar symmetry in the dependence of the deformation
of a vertex operator in an orbifold model on the conformal dimension
about the point $h=1/8$, observed in \cite{Ebe01}.



\vspace{0.7cm}
\noindent
{\bf Acknowledgements.}
I am very grateful to Werner Nahm for his guidance and very helpful discussions
as well as to Katrin Wendland and Michael Flohr.
I would like to thank the Studienstiftung des Deutschen Volkes for support and
the Dublin Institute of Advanced Studies for hospitality.


\begin{appendix}
\section{Explicit identification of lattice vectors\label{appA}}
In order to prove the theorem in section \ref{ident} we have explicitely checked
the identification for all $42$ different generators of $\Gamma({\cal K} (\Z^4,0))$ 
described in the equations \mref{Z2_eq1} and \mref{Z2_eq2} (with lattice $\Lambda = \Z^4$).
In this appendix we collect a variety of typical examples of this calculation.
Each time we first express a generator of $\Gamma({\cal K} (\Z^4,0))$ in terms of 
the orthogonal basis of $A_i$, then we use the identification given in section \ref{ident}
to rewrite this expression in terms of the $B_j$, and finally we express
this in terms of a sum of lattice vectors of 
$\Gamma(\Z_4 \frac{1}{\sqrt{2}} D_4, B^*)$:
\beq
\hat{E}_{0000} &=& \frac{1}{4} (A_{5}+A_{6}+A_{7}+A_{8}+A_{13}+A_{14}+A_{15}+A_{16}+A_{4}-A_{24}) \nonumber \\
&\widehat{=}& \frac{1}{4} (B_{2}-B_{13}+B_{14}+B_{16}-B_{9}-B_{10}+B_{11}+B_{12}-B_{15}-B_{24}) \nonumber \\
&=& - \hat{\upsilon} - \hat{E}_{1100}^{(+)} - \hat{E}_{1100}^{(0)} - \hat{E}_{1100}^{(-)} \nonumber \\
\hat{E}_{1000} &=& \frac{1}{4} (A_{9}+A_{10}+A_{11}+A_{12}+A_{17}+A_{18}+A_{19}+A_{20}+A_{4}-A_{24}) \nonumber \\
&\widehat{=}& \frac{1}{4} (-B_{1}+B_{3}+B_{8}+B_{7}-B_{17}-B_{18}+B_{19}-B_{20}-B_{15}-B_{24}) \nonumber \\
&=& \left(-e_1 \wedge e_2 + \frac{1}{2} (\hat{E}_{0000}^{(+)} + \hat{E}_{0000}^{(-)} + 
\hat{E}_{0011}^{(+)} + \hat{E}_{0011}^{(-)}) \right)  \nonumber \\
&& + \left(\frac{1}{2} e_3 \wedge e_4 -\frac{1}{2} \hat{E}_{0100}
- \frac{1}{4} (-\hat{E}_{0000}^{(+)} - 2\hat{E}_{0000}^{(0)} - 3\hat{E}_{0000}^{(-)} 
- \hat{E}_{1100}^{(+)} - 2\hat{E}_{1100}^{(0)} - 3\hat{E}_{1100}^{(-)}) \right) \nonumber \\
&& + \left(-\frac{1}{2} (e_1 \wedge e_3 + e_4 \wedge e_2 ) - \frac{1}{2} (\hat{E}_{0000}^{(+)} + \hat{E}_{0000}^{(-)})
+ \frac{1}{2} (\hat{E}_{0100} + \hat{E}_{0001} + \hat{E}_{0101}) \right) \nonumber \\
&& + \left(\frac{1}{2} (e_1 \wedge e_4 + e_2 \wedge e_3 ) + \frac{1}{2} (\hat{E}_{0011}^{(+)} + \hat{E}_{0011}^{(-)})
+ \frac{1}{2} (- \hat{E}_{0111} - \hat{E}_{0001} - \hat{E}_{0101}) \right) \nonumber \\
&& - \hat{\upsilon}^{\circ}  - \hat{E}_{0000}^{(+)} - \hat{E}_{0000}^{(0)} - \hat{E}_{0000}^{(-)} 
-  \hat{E}_{1100}^{(+)} - \hat{E}_{1100}^{(0)} - \hat{E}_{1100}^{(-)} 
- \hat{E}_{0011}^{(+)} - \hat{E}_{0011}^{(-)}\nonumber \\
\lefteqn{\!\!\!\!\!\!\!\!\!\!\!\!\!\!\!\! \frac{1}{\sqrt{2}} e_1 \wedge e_2 + 
\frac{1}{2} (\hat{E}_{0000} - \hat{E}_{0010} - \hat{E}_{0001} + \hat{E}_{0011})} \nonumber \\
&=& \frac{1}{4} (A_{1}+A_{21}+A_{7}+A_{8}+A_{13}+A_{14}-A_{19}+A_{20}-A_{9}+A_{10}) \nonumber \\
&\widehat{=}& \frac{1}{4} (B_{6}-B_{21}+B_{14}+B_{16}-B_{9}-B_{10}-B_{19}-B_{20}+B_{1}+B_{3}) \nonumber \\
&=& \left(- \frac{1}{2} e_3 \wedge e_4\ + \frac{1}{2} \hat{E}_{0100} + \frac{1}{4} (-\hat{E}_{0000}^{(+)} 
- 2\hat{E}_{0000}^{(0)} - 3\hat{E}_{0000}^{(-)} - \hat{E}_{1100}^{(+)} 
- 2\hat{E}_{1100}^{(0)} - 3\hat{E}_{1100}^{(-)}) \right) \nonumber \\
&& \quad + \hat{E}_{0000}^{(-)} \nonumber \\
\hat{\upsilon} &=& \frac{1}{2} (-A_{4} + A_{24}) \cong \frac{1}{2} (B_{15} + B_{24}) \nonumber \\
&\widehat{=}& \left(e_1 \wedge e_2 +   \frac{1}{2} (\hat{E}_{0000}^{(+)} + \hat{E}_{0000}^{(-)} + 
\hat{E}_{0011}^{(+)} + \hat{E}_{0011}^{(-)}) \right)  \nonumber \\
&& + \left(\frac{1}{2} (e_1 \wedge e_3 + e_4 \wedge e_2 ) - \frac{1}{2} (\hat{E}_{0000}^{(+)} + \hat{E}_{0000}^{(-)})
+ \frac{1}{2} (\hat{E}_{0100} + \hat{E}_{0001} + \hat{E}_{0101}) \right) \nonumber \\
&& + \left(- \frac{1}{2} (e_1 \wedge e_4 + e_2 \wedge e_3 ) + \frac{1}{2} (\hat{E}_{1111}^{(+)} + \hat{E}_{1111}^{(-)})
+ \frac{1}{2} (\hat{E}_{0111} + \hat{E}_{1110} + \hat{E}_{0110}) \right) \nonumber \\
&& + 2 \hat{\upsilon}^{\circ} + 2 \hat{\upsilon} + \sum_{i \in \{0000,1100\}} (\hat{E}_{i}^{(+)} + 
\hat{E}_{i}^{(0)} + \hat{E}_{i}^{(0)} ) +  \hat{E}_{0011}^{(0)} + \hat{E}_{1111}^{(0)} \nonumber \\
\hat{\upsilon}^{\circ} &=& \frac{1}{4} (-A_{4}+A_{5}-A_{6}-A_{7}-A_{8}-A_{17}-A_{18}-A_{19}+A_{20}+3 A_{24}) \nonumber \\
&\widehat{=}& \frac{1}{4} (B_{15}+B_{2}+B_{13}-B_{14}-B_{16}+B_{17}+B_{18}-B_{19}-B_{20}+3 B_{24}) \nonumber \\
&=& 4 \hat{\upsilon}^{\circ} + 4 \hat{\upsilon} +2 e_1 \wedge e_2 + (e_1 \wedge e_3 + e_4 \wedge e_2 ) 
- (e_1 \wedge e_4 + e_2 \wedge e_3 ) + \sum_{i \in I^{(2)}} \hat{E}_{i} \nonumber \\
&& + 2 \sum _{i \in I^{(4)}} \left(\hat{E}_{i}^{(+)} + \hat{E}_{i}^{(0)} + \hat{E}_{i}^{(-)} \right)
- \frac{1}{2} \sum _{i \in I^{(4)}} \left(\hat{E}_{i}^{(+)} + \hat{E}_{i}^{(-)} \right) + \hat{E}_{1100}^{(+)} \; .
\eeq

\section{R-sector of the $\Z_4$ orbifold partition function\label{Rsector}}
Using projection and modular transformations one can calculate the partition function
of a $\Z_4$ orbifold from its original torus theory, always keeping the characters
of the bosonic and the fermionic part separate factors. We are only interested in the
R-sector of this partition function
\beq
Z_{\Z_4}^{\Gamma, \, R} &=& \frac{1}{4} \left( \left( \frac{1}{|\eta(\sigma)|^8} \sum_{p\in \Gamma} q^{p_l^2/2}
\qb^{p_r^2/2} \right) * \left| \frac{\theta_2(\sigma,z)}{\eta(\sigma)} \right|^4
+ 8 \left| \frac{\eta(2\sigma)}{\theta_2(2\sigma)} \right|^2 * \left| \frac{\theta_2(2\sigma,2z)}{\eta(2\sigma)} \right|^2
\right. \nonumber \\
&& + 16 \left| \frac{\eta(\sigma)}{\theta_2(\sigma)} \right|^4 * \left| \frac{\theta_1(\sigma,z)}{\eta(\sigma)} \right|^4
+ 16 \left| \frac{\eta(\sigma)}{\theta_4(\sigma)} \right|^4 * \left| \frac{\theta_3(\sigma,z)}{\eta(\sigma)} \right|^4
+ 16 \left| \frac{\eta(\sigma)}{\theta_3(\sigma)} \right|^4 * \left| \frac{\theta_4(\sigma,z)}{\eta(\sigma)} \right|^4
\nonumber \\
&& + 8 \left| \frac{\eta(\frac{\sigma}{2})}{\theta_4(\frac{\sigma}{2})} \right|^4 * 
\left| \frac{\theta_3(\frac{\sigma}{2},z)}{\eta(\frac{\sigma}{2})} \right|^4
+ 8 \left| \frac{(\eta(\sigma) \theta_3(\sigma))^{1/2}}{\theta_3(2\sigma) - i \theta_2(2\sigma)} \right|^2 
* \left| \frac{(\theta_3(2\sigma,2z) + i \theta_2(2\sigma,2z))}{(\eta(\sigma) \theta_3(\sigma))^{1/2}} \right|^2 \nonumber \\
&& + 8 \left| \frac{(\eta(\sigma) \theta_4(\sigma))^{1/2}}{\theta_3(2\sigma) + \theta_2(2\sigma)} \right|^2 
* \left| \frac{(\theta_3(2\sigma,2z) - \theta_2(2\sigma,2z))}{(\eta(\sigma) \theta_4(\sigma))^{1/2}} \right|^2 \nonumber \\
&& + 8 \left| \frac{(\eta(\sigma) \theta_3(\sigma))^{1/2}}{\theta_3(2\sigma) + i \theta_2(2\sigma)} \right|^2 
* \left| \frac{(\theta_3(2\sigma,2z) - i \theta_2(2\sigma,2z))}{(\eta(\sigma) \theta_3(\sigma))^{1/2}} \right|^2 \nonumber \\
&& \left .+ 8 \left| \frac{(\eta(\sigma) \theta_2(\sigma))^{1/2}}{\theta_3(\frac{\sigma}{2}) 
+ \theta_4(\frac{\sigma}{2})} \right|^2 
* \left| \frac{(\theta_3(\frac{\sigma}{2},2z) + 
\theta_4(\frac{\sigma}{2},2z))}{(\eta(\sigma) \theta_2(\sigma))^{1/2}} \right|^2 \right) \; ;
\eeq
in each term the first factor gives the bosonic part, the second the fermionic. The first three terms constitute
the untwisted sector, the other seven terms the different twisted sectors. Expanding the terms
six to nine yields a leading contribution of
\beq
2 \left( (q\qb)^{-1/6} \, (q\qb)^{3/16} \right) \, * \, \left( (q\qb)^{-1/12} \, (q\qb)^{1/16} \right) \, * 1
\eeq
for each. The first big bracket gives the bosonic contribution, separating the overall modular factor first, 
the second bracket gives the fermionic contribution. Hence we find eight fields of overall conformal weight
$(h,\hb)=(\frac{1}{4}, \frac{1}{4})$ where a $h_b = \frac{3}{16}$ part originates from bosonic degrees of freedom, 
a $h_f = \frac{1}{16}$ part from fermionic.
This perfectly well coincides with the conformal weights of twistfields generating cuts for either bosonic
or fermionic fields found in \cite{Dix87}. 

\end{appendix}


\bibliographystyle{JHEP}
\bibliography{holgerJHEP}

\end{document}